# Score function for the optimization of the performance of forward fill/ flush differential flow modulation for comprehensive two-dimensional gas chromatography

Aleksandra Lelevic[a,b,*], Christophe Geantet[b], Chantal Lorentz[b], Maxime Moreaud[a], Vincent Souchon[a]

a. IFP Energies nouvelles, Rond-point de l'échangeur de Solaize BP 3 69360 Solaize France
b. Univ Lyon, Université Claude Bernard Lyon 1, CNRS, IRCELYON, F-69626, Villeurbanne, France

* Author for correspondence: aleksandra.lelevic@ifpen.fr

**Abbreviations**

$^1$D - First dimension

$^2$D - Second dimension

CFT - Capillary flow technology

FID - Flame ionisation detector

GC×GC - Comprehensive two-dimensional gas chromatography

MS - Mass spectrometry

PLS - Partial Least Squares

## Abstract

Modulation is the key element of the comprehensive two-dimensional gas chromatography separation. Forward fill/flush flow modulation is cost effective, robust and suitable for analysis of a wide range of samples. Even though this modulation system is well known, studies regarding its optimization are sparse. In this work, based on hundreds of experiments involving multiple column sets and modulation conditions, an approach was proposed that permits to facilitate the choice of the forward fill/flush flow modulation parameters. A score function was developed that allows to predict the forward fill/flush flow modulation process efficiency as judged by the modulated peak shape. The score function was based on the physical rules for optimized and quantitative forward fill/flush flow modulation proposed in our previous work which state that the sum of the fill and flush modulation distance should be close to the modulation channel length and that the ratio of the flush and fill distance should be sufficiently high for efficient channel flushing. The score function was embedded in a freely available tool in the form of a forward fill/flush flow modulation calculator which allows the user to quickly check the relevancy of the modulation operating conditions or to obtain a suggestion for optimal modulation parameters.

## 1. Introduction

Modulator is a crucial component of the GC×GC set-up. It is often regarded as the bottleneck element of the GC×GC system especially in systems involving flow-modulation, owing to the technical limitations of its performance. Modulators are often classified into three categories: thermal, valve-based and flow modulators [1–3].

In general, cryogenic modulators provide higher peak capacity and sensitivity, they involve easier method development and have no limitations when it comes to coupling to MS due to compatible carrier gas flow rates. On the other hand, flow modulators are characterized by low cost and robustness, no use of cryogen and no volatility limits with respect to GC analysis, which makes them a very good choice for routine analysis of a large variety of samples. Flow modulation for GC×GC has been first introduced by Seeley et al. in 2000 [4] with a further refinement of the technique by Bueno and Seeley [5] which involved the elimination of the diaphragm valve and allowed 100% mass transfer. This development led to commercialization of the technique and introduction of the capillary flow technology (CFT) modulator by Agilent which is studied in this work. This type of modulator has been investigated and compared to cryo modulation in the work by Semard et al. [6], who found that the differential flow modulation can result in a similar peak capacity as cryo modulation in the optimized conditions.

The use of forward fill/flush flow modulator has increased in the last years, covering many types of applications, as illustrated in the Table I. However, operation of the flow modulator is more difficult to optimize due to hardware restrictions and small number of tuneable parameters. If not carefully optimized, issues with modulated peak shape and compromised quantitative performance might result [7]. Several studies in the past have addressed the issue of improved performance of the flow modulation for certain applications. Some developments have been proposed, such as the use of pulsed flow modulation by Amirav et al. [8,9]. Dynamic pressure gradient modulation generating very narrow peaks with very low modulation periods was proposed by Trinklein et al. [10]. Tranchida et al. [11] have investigated flow modulation low pressure GC×GC by employing mega-bore second-dimension column and a long accumulation loop which resulted in improved peak shapes and low $^{2}$D flows compatible with MS analysis. The use of reverse fill/flush flow modulation has been proposed by Griffith et al. [12], which reportedly can result in improved peak shapes and sensitivity. For this type of modulation system, Giardina et al. [13] have proposed a pneumatic model for its easier optimisation.

Regarding forward fill/flush flow modulator however, it is not well documented how parameters of the modulation process ought to be selected and what is the impact of varying individual parameters on the modulation efficiency. In our previous work [7], we have discussed the main factors that influence the efficiency of the modulation process. For a fixed GC×GC system, these were identified to be: $^{1}$D column flow, $^{2}$D column flow, and modulator sampling and injection times (modulation period = sampling time + injection time). These parameters determine the traversed distances of the GC effluent in the modulation channel during the two modulation stages (fill and flush) which are of crucial importance for the performance of modulation. The efficiency of the flow modulation was judged based on generated modulated peak shapes, which proved to be in the direct connection with the quantitative performance of the modulation. It was observed that good modulated peak shape characterized by low tailing and/or fronting with good analyte quantification was obtained when the sum of the modulation fill and flush distances was ~15 cm which was the length of the modulation channel with a sufficiently high ratio of flush and fill distances (> ~2.5) ensuring good flushing of the modulation channel. The tolerance for the deviation of the sum of the fill and flush distance from the channel length was determined to be ±2 cm, beyond this value serios quantification and peak shape issues were incurred. Additionally, it was found that in a thermally programmed GC×GC run owing to the increase of oven temperature, concomitant carrier gas velocity changes lead to fine changes of the distances travelled during the two steps of the modulation process changing its efficiency progressively. Thus, different performances of modulation from the beginning to the end of thermally programmed GC×GC analysis were obtained, which can lead to quantitative issues especially when analysing samples containing analytes with a wide range of molecular masses.

Based on these previously established rules for efficient modulation, we have tried in this work to expand this previously developed approach and to predict modulation process efficiency for any set of chosen modulation parameters. Attempts to predict modulation efficiency for a reverse fill-flush flow modulation system have been made by LECO [28] in their ChromaTOF software where warning is issued if for example sampling loop is overfilled or calculated volume flush factor is not suitable. To achieve a prediction of performance for forward fill/flush flow modulation system, we first expanded the scope of our previous work to different column sets, oven programming and carrier gases, in order to verify if the generic rules for efficient modulation established in our previous study can be well applied in a wider range of GC conditions. Based on these generic rules a score function was developed which allows to grade modulation process efficiency and thus predict modulation performance for selected conditions. This hypothesis has been tested by investigating modulated peak shapes and calculating score function for hundreds of GC×GC runs which have been performed in a wide range of GC conditions. Finally, this score function was embedded in a tool in the form of a forward fill/flush flow modulation calculator which enables the user to quickly check the expected performance of modulation for a chosen set of separation parameters or obtain a suggestion for modulation conditions which best suit the selected separation criteria. This work is thus a first step regarding our approach for optimization of flow modulation and will be further expanded to include more complex systems involving secondary column temperature programming, or other types of modulation system such as reversed fill/flush. A deeper insight in general is needed also into the influence of the choice of modulation parameters in these systems on the chromatographic separation which needs to be further investigated but is not a subject of the present work.

## 2. Experimental

For this study, a standard mixture of n-paraffins from n-$C_{10}$ to n-$C_{28}$ (n-$C_{10}$, n-$C_{12}$, n-$C_{16}$, n-$C_{18}$, n-$C_{20}$, n-$C_{22}$, n-$C_{24}$, n-$C_{26}$ and n-$C_{28}$) diluted in toluene was used for optimization and quantitative performance tests. For High Temperature (HT) GC×GC analysis, a standard mixture of n-paraffins from n-$C_{10}$ to n-$C_{44}$ (n-$C_{10}$, n-$C_{12}$, n-$C_{15}$, n-$C_{16}$, n-$C_{18}$, n-$C_{20}$, n-$C_{22}$, n-$C_{24}$, n-$C_{26}$, n-$C_{28}$, n-$C_{30}$, n-$C_{32}$, n-$C_{36}$, n-$C_{38}$, n-$C_{44}$) diluted in carbon disulphide and toluene was used. All chemicals used to prepare the n-paraffin mixture were of analytical grade quality (Sigma–Aldrich). Concentrations of the test mixtures are provided in the Supplementary material. Additionally, 00.02.718 PNA in Diesel - Gravimetric blend from AC Analytical Controls® (PAC) (n-paraffins, naphthenes, FAMEs, mono and diaromatics) was analysed in order to demonstrate the performance of flow modulation calculator for a more complex test mixture.

Agilent 7890A gas chromatograph was employed, equipped with a G3486A CFT differential flow modulator and FID detection. Our investigation was limited to "normal-phase" column configurations that are commonly applied (Table I). Several column sets were tested. I: DB-1 (100% dimethylpolysiloxane 20 m, 0.1 mm ID, 0.4 μm) × BPX-50 (50% phenyl polysilphenylene-siloxane 3.2 m, 0.25 mm ID, 0.25 μm), II: DB-1 (100% dimethylpolysiloxane 20 m, 0.1 mm ID, 0.4 μm) × BPX-50 (50% phenyl polysilphenylene-siloxane 5 m, 0.25 mm ID, 0.25 μm), III: Rxi-1ms (100% dimethylpolysiloxane 30 m, 0.25 mm ID, 0.25 μm) × ZB-35HT (35% phenyl 65% dimethylpolysiloxane 5 m, 0.25 mm ID, 0.18 μm), IV: ZB-5HT (5%-phenyl 95% dimethylpolysiloxane 15 m, 0.1 mm ID, 0.1 μm) × ZB-35HT (35% phenyl 65% dimethylpolysiloxane 5 m, 0.25 mm ID, 0.18 μm). Carrier gas was hydrogen or helium. 1 μL injections with a split ratio of 150:1 were performed on a MMI Agilent inlet equipped with a single taper liner with glass wool. Injection port was heated to 300 °C, then ramped to 330 °C at 500 °C/min, where it remained isothermal during 5 min. Oven temperature program was: 50 °C (3 min) –325 °C, ramps 1.5 °C/min, 2.5 °C/min and 3 °C/min were tested. FID conditions were as follows: 325 °C, air flow 400 ml/min, hydrogen was changed depending on $^2$D flow so that hydrogen flow + $^2$D flow is ca. 10% of air flow, and make-up gas (nitrogen) 25 ml/min. For HT GC analysis (column set IV), injection port was heated to 300 °C, then ramped to 400 °C at 500 °C/min, where it remained isothermal during 5 min. Oven temperature program was: 60 °C (3 min) – 390 °C (2.5 °C/min). FID conditions were as follows: 360 °C, air flow 400 ml/min, hydrogen 20 ml/min and make-up gas (nitrogen) 25 ml/min. Throughout the analysis campaign different modulation conditions were tested, involving various $^1D$ and $^2D$ column flows, modulation periods and injection times.

Agilent ChemStation B.04.03-SP1 was used for GC data acquisition. Python 3.7 was employed for designing flow modulation calculator and modulated peak shape investigation. PLS (Partial Least Squares) analysis was performed in OriginPro 2019.

## 3. Results and discussion

### 3.1. An approach for investigating modulated peak shape

Good modulated peak shape is essential in ensuring flow modulation process efficiency [8,22,29–32]. Modulated peak shape with minimal tailing and fronting that returns to the baseline before the next modulation cycle ensures good peak integration and thus it is directly connected with analyte quantification as also demonstrated in our previous work [7]. For the investigation of the modulated peak shape, it is difficult to apply conventional metrics, such as for example asymmetry factor or USP (U.S. Pharmacopeia) tailing factor. This is because an improperly modulated peak often does not return to the baseline between individual modulations, thus it is often difficult to determine peak start and its end, leading to incorrect peak integration. Additionally concurrent tailing and fronting, possibly even double peaks, can often occur which are difficult to characterize with single number metrics. And in the end, for proper investigation it is necessary to look at a modulated peak as a whole and not only individual modulation peaks.

Thus, for investigating a global shape of any modulated peak, we have developed a simple approach in which we fit gaussian peaks below each individual measured peak and we assume that in ideal case signal should return to the zero baseline between the fitted peaks. We have chosen this approach due to its simplicity, but also as our goal was not to model "real" peak shapes but perform simple comparison between peak shapes issued from different modulation conditions. Gaussian peak retention time, height and standard deviation were optimized so as to obtain minimum residuals for the fit with the upper half of the measured peak. For example, in Figure 1A a measured modulated peak of n-$C_{10}$ alkane, which demonstrates global tailing symptomatic of improper modulation channel flushing, is traced in black. To each of the four modulated peaks a gaussian peak was fitted, as shown in red. Then the difference of the measured signal and the fitted signal was calculated-traced in dashed blue line. Peak shape is often not perfect in gas chromatography and small tailing and/or fronting of the peak that returns to the baseline is not detrimental for proper quantification. Bigger issue in peak modulation is the lack of the peak return to the baseline between individual modulation cycles which prevents proper peak integration. Proposed approach particularly puts into perspective such modulated peak behaviour. For comparison of the modulated peak shapes from different analysis, normalization of the signal was performed by dividing each modulated peak signal by its measured total area, as peak area is directly proportional to the absolute mass of the analyte. Integrated area for a modulated peak is illustrated in Figure S1 in the Supplementary material. Thus, obtained residuals for each modulated peak were scaled to the same unit mass and could have been directly compared. In this way, the best modulation performance was defined as the one that results in minimal residuals between the measured and "ideal" peak shape. This approach was used to evaluate and compare modulation process efficiency all along our study.

### 3.2. Influence of modulation parameters: theoretical calculations

To study how changes in operating GC×GC conditions ($^{1}$D column flow, $^{2}$D column flow, modulator sampling and injection times, $^{1}$D and $^{2}$D column length and diameter) influence modulation performance, theoretical calculations were first performed. Modulator channel length was not considered, as in this system modulation channel length is fixed. To simulate a large set of GC×GC conditions, possible values of $^{1}$D column dimensions were chosen to be 15, 20 and 25 m for length, 0.25 and 0.1 mm for internal diameter. For $^{2}$D column, lengths of 3, 4 and 5 m and internal diameters of 0.32 and 0.25 mm were looked at. $^{1}$D column flows of 0.15, 0.2, 0.3 and 0.6 ml/min and $^{2}$D column flows of 6, 10, 15, 20 and 25 ml/min were chosen. Modulation periods of 1, 3, 4, 6 and 8 s and injection times of 0.12, 0.14, 0.16, 0.18, 0.20, 0.22 and 0.25 s were selected. Hydrogen was chosen as a carrier gas. According to system of equations already presented in [7], fill and flush distances were calculated for all possible combinations of these parameters at four different temperatures 70 °C, 155 °C, 240 °C and 325 °C, simulating a classical temperature-programmed separation.

PLS (Partial Least Squares) analysis was performed on these data to study the dependency of four variables regarding operating parameters. These 4 dependent variables were taken to be: fill [cm] and flush distances [cm] at initial analysis temperature (70 °C), difference of the sum of fill and flush distances at the analysis final (325 °C) and initial temperature (70 °C) and the ratio of flush to fill distance (which is the same at all temperatures). Results of the PLS analysis are presented in Figure 2. Plotted vectors for each dependent variable (in blue) give some information on how these are correlated to operating parameters (in red): parallel vectors indicate that they are positively correlated, anti-parallel vectors are obtained for negatively correlated variables whereas orthogonal vectors are characteristic of independent variables and parameters.

Fill distance mostly depends on $^{1}$D flow and modulation period (more specifically sampling time) as expected. It is inversely related to $^{2}$D column flow as the increase of the modulator outlet pressure (equal to $^{2}$D column head pressure) leads to decrease of average velocity in the modulator channel. For the same reason, fill distance is affected by $^{2}$D column diameter and length as increase of $^{2}$D column length or decrease of $^{2}$D column diameter both lead to increase of $^{2}$D column head pressure. Flush distance is on the other hand mostly affected by $^{2}$D flow and injection time and does not depend on $^{1}$D column flow or modulator sampling time. Increasing $^{2}$D column length or decreasing its diameter leads also in this case to decrease of velocity leading to decrease of flush distance. Changing $^{1}$D column dimensions affects neither fill nor flush distance, nor consequently other chosen descriptors of the modulation process efficiency.

Ratio of the fill and flush distances is affected by the four parameters of the modulation process: $^{1}$D column flow, $^{2}$D column flow, modulator sampling and injection time, however it does not seem to depend on chosen column geometry on both first and second dimensions.

Most interestingly perhaps, change of the sum of the fill and flush distances from the initial to the final analysis temperature is mostly affected by $^{2}$D column dimensions. As $^{2}$D column length is increased or its diameter decreased, the increase of its head pressure leads to increase of modulator outlet pressure, which in turn decreases the change of average velocity in the modulator from the beginning to the end of the run. As any changes of modulator performance during the run ought to be minimized, when using flow modulation it is better to work with longer $^{2}$D columns or smaller $^{2}$D diameters. However, long columns with small diameters may not always be suitable, as they involve high pressures, long void times, increased retention, thus compromise should be made between efficient separation and efficient modulation.

Additionally to PLS visualization, Pearson correlation coefficients were calculated for all variables. Results are displayed in Figure S2 in the Supplementary material and illustrate in the same way the above-described dependences.

### 3.3. Defining a score function for modulation performance evaluation

The purpose of this work was to define a score function that can grade the modulation process on the basis of the above-mentioned performance criteria. For the function, specifically considered were following criteria: sum of fill and flush distance at initial analysis temperature, sum of fill and flush distance at final analysis temperature and the ratio of flush to fill distance (which was the same at all temperatures).

Both initial and final analysis temperatures were considered because in the case of a sample with compounds having a wide range of volatility, modulated peaks might exhibit improved or deteriorated modulation process efficiency from the start to the end of the run. This is due to the fact that, as oven temperature rises in the thermally programmed run average velocities in the modulation channel increase for the same parameters of the modulation process, leading to new values for the fill and flush distances for every subsequent peak. To take into account these changes in the sum of fill and flush distances with temperature and ensure proper modulation across the entire GC run, it was decided to introduce in the score function the corresponding values for the analysis initial and final temperature.

As demonstrated in our previous work [7] for the investigated type of modulator for a given GC×GC column set, sum of the fill and flush distances should be close to the length of the modulation channel and the ratio of the flush and fill distance should be high enough to ensure proper channel flushing. Thus, these results were taken as target values for the modulation process performance criteria.

Finally, the score function $F_{(L)}$ was then defined in the following way:

$$F_{(L)} = 2 \cdot (E_1 + E_2) + E_3 \tag{1}$$

$$E_1 = \begin{cases} |L^{T_{in}} - 15|, & if\ L^{T_{in}} \in [12.3, 15.5] \\ 3 \cdot |L^{T_{in}} - 15|, & if\ L^{T_{in}} < 11.5\ or\ L^{T_{in}} > 16 \\ 1.5 \cdot |L^{T_{in}} - 15|, & else \end{cases} \tag{2}$$

$$E_2 = \begin{cases} \left|L^{T_{fin}} - 15\right|, & if\ 1L^{T_{fin}} \in [14.5,17.5] \\ 3 \cdot \left|L^{T_{fin}} - 15\right|, & if\ L^{T_{fin}} < 14\ or\ L^{T_{fin}} > 21 \\ 1.5 \cdot \left|L^{T_{fin}} - 15\right|, & else \end{cases} \tag{3}$$

$$E_3 = \begin{cases} 5, & if\ \frac{L_{flush}^{T_{fin}}}{L_{fill}^{T_{fin}}} < 1 \\ -\frac{\left(\frac{L_{flush}^{T_{fin}}}{L_{fill}^{T_{fin}}} - 1\right)}{2}, & if\ \frac{L_{flush}^{T_{fin}}}{L_{fill}^{T_{fin}}} \in [1,5] \\ -2.5, & else \end{cases} \tag{4}$$

where $L^{T_{ini}}$ and $L^{T_{fin}}$ are the sum of fill and flush distances at analysis initial and final temperature, respectively. $L_{flush}^{T_{fin}}$ is the flush distance and $L_{fill}^{T_{fin}}$ fill distance at analysis final temperature.

Developed score function was based on a system of penalty or reward points depending on the traversed fill and flush distances at different temperatures and their ratio. The $E_1$ term of the score function was dedicated to sum of fill and flush distances at analysis initial temperature, awarding different number of points, depending how close this sum is to the assumed length of the modulation channel i.e. ~15 cm. Smaller penalty is awarded for small deviations, and it is increasing when going further from optimal length (three levels of deviations are defined based on our previous work [7]). $E_2$ is a similar term, however this element is related to the sum of fill and flush distances at analysis final temperature. Term $E_3$ is dedicated to the ratio of the flush and fill distances, where highest penalty is applied if the ratio of the flush and fill distances is smaller than 1, which means that the flush distance would be smaller then fill distance.

Overview of the calculated function values for all possible combination of conditions mentioned in previous section (25 200 combinations) is illustrated in Figure 3. The value of the score function can range from ~1 up to high values. The higher the value, the poorer the modulation performance as the higher score function value means more penalty points awarded due to furthering away from optimal performance criteria.

In the case where all the modulation parameters are set except one for example, testing all possible values of this parameter and choosing the one that results in a minimal value of the $F_{(L)}$ score function leads to the best predicted performance of the modulation with respect to set target conditions. For example, this can be done to optimize modulation injection times, or $^2$D Flows but more generally for every modulation parameter provided that the other ones are defined.

### 3.4. Optimizing modulation parameters with a score function

For the validation of the defined score function, a large set of experiments was performed, where four different GC×GC set-ups were employed. Both $H_2$ and He were tested as carrier gases, and oven ramp was varied from 1.5 to 3 °C/min in order to ensure good estimation for a range of commonly employed conditions. More than 200 runs were performed in which a paraffin test mixture was analysed in various combinations of $^1$D flow, $^2$D flow, modulation period and modulation injection time. A real sample was not used for the tests, as complex matrices often do not permit to reliably investigate modulated peak shape due to every modulation normally containing multiple peaks present. However, a commercial gravimetric blend containing n-paraffins, naphthenes, FAMEs, mono and diaromatics was analysed in order to test the performance of the score function for a more complex test mixture. Overview of the selected separation conditions presented in this experimental section is provided in the Table II.

Criterium for determination of the modulation process efficiency was looking at modulated peak shape across the chromatogram which was shown to be directly related to the quantification performance in our previous work [7] and thus to the modulation efficiency. Modulated peak shape was investigated as described in the Section 3.1.

#### 3.4.1. Optimizing modulation injection time with the score function

A first set of analysis was performed under hydrogen as carrier gas with column set I, *id est* DB-1 (20 m, 0.1 mm ID, 0.4 µm) and BPX-50 (3.2 m, 0.25 mm ID, 0.25 µm) columns. The working values of $^1$D flow, $^2$D flow and modulation period are generally chosen so as to optimize the GC×GC separation. Hence, the question that most often poses itself is which modulator injection time is suitable for the selected conditions. To demonstrate how the result of the defined score function can offer useful guidance for choosing the most appropriate modulation injection time, the following conditions were set: chosen $^1$D flow was 0.15 ml/min, $^2$D flow was 11.7 ml/min and modulation period was 4.5 s. $F_{(L)}$ values were calculated for injection times from 0.12 to 0.23 s and these results along with the accompanying values of fill and flush distances are provided in Table III. Initial and final temperature were taken to be close to elution temperatures for the lightest and heaviest n-paraffin in the test mixture.

The lowest values for the score function were obtained for injection times around 0.14-0.16 s as the sum of flush and fill distance and their ratio were deviating the least from the optimal values. If the injection time is lower than 0.14-0.16 s, according to calculated flush and fill distances, under-flushing of the modulation channel ought to be observed, and in the same way much higher injection times ought to cause over-flushing of the channel and breakthrough of the $^1$D effluent.

The associated measured chromatograms confirm that 0.14-0.16 s injection times give the best results. Figures 4A and 4B show modulated peaks of n-$C_{10}$ and n-$C_{28}$, respectively for the runs identified as the best. Peak shapes are judged to be acceptable, increasing injection time starts to bring about small fronting for the peak of n-$C_{28}$ as the sum of fill and flush distances approaches high values (17 cm for 0.16 s). In Figures 4C and 4D, chromatograms obtained for 0.12, 0.15 and 0.18 s injection times are compared. For 0.12 s, tailing for n-$C_{10}$ peak can be seen in Figure 4C in the insert - trace in blue in line with low fill+flush distance 11.69 cm. Being too high, 0.18 s causes fronting for heavy analytes which is prominent for the n-$C_{28}$ peak as it can be seen in Figure 4D in the insert - trace in red, where double peaks start to appear (fill+flush 18.55 cm). Even higher injection times, as for example 0.23 s, lead to a very high $F_{(L)}$ value indicating serious peak shape issues, which is corroborated by the chromatogram obtained, where double peaks are both seen at the beginning (n-$C_{10}$) and at the end of the chromatogram (n-$C_{28}$), as depicted on Figures 4E and 4F. Normalized residuals from fitted gaussian signal are shown in Figure S3 in the Supplementary material for every chromatogram obtained for injection time 0.12-0.18 s. As seen in the original chromatograms, 0.12 s causes dominant tailing seen for the n-$C_{10}$ peak, while 0.18 s causes peak fronting for heaviest peaks, here demonstrated for n-$C_{28}$. Thus, the best compromise, also as predicted by the score function is found between these values.

### 3.4.2. Optimizing $^2$D flow with the score function

Similar results are obtained when looking for an optimal modulation period or $^2$D flow. To demonstrate the example of choosing the suitable $^2$D flow by minimizing the score function, the following conditions were chosen: $^1$D flow 0.1 ml/min, modulation period 5.5 s and injection time 0.18 s. Results obtained for the $F_{(L)}$ with $^2$D flows from 6 to 18 ml/min are provided in Table IV.

In these conditions, $^2$D flows around 10 ml/min seem to be suitable as they lead to the lowest values for the score function. Lower flows as in the case of injection times cause under-flushing, and higher lead to over-flushing of the modulation channel. This is supported also by the acquired chromatograms. Figures 5C and 5D show modulated peaks for n-$C_{10}$ and n-$C_{28}$, for 10 mL/min $^2$D flow. Good peak shape testifies satisfactory modulation performance. Figures 5A and 5B show modulated peaks for a 6 ml/min $^2$D flow. For n-$C_{10}$, peak tailing is prominent due to insufficient flushing, sum of fill and flush distances being only around 11 cm. This sum increases as oven temperature rises, and in the end tailing is much less prominent as it can be seen for the n-$C_{28}$ peak (fill+flush 13.77 cm). Due to the tailing which however dominates for earlier peaks, modulation performance is compromised. Opposite issue is obtained with $^2$D flow of 14 ml/min. In this case, peaks start to demonstrate fronting, which is even more severe for heavier analytes and further exacerbated by even higher $^2$D flows. Normalized residuals from fitted gaussian signal are shown in Figure S4 in the Supplementary material for every chromatogram obtained for $^2$D Flows 6-14 ml/min. Trends observed on the original chromatograms are corroborated by peak shapes investigation. Low $^2$D Flow of 6 ml/min causes dominant tailing perceived for light peaks, while at 14 ml/min peak fronting is perceived. It can be concluded that also in this case, the score function gives a good assessment of suitable conditions of $^2$D flow.

### 3.4.3. Checking the applicability of the score function to other conditions

With the same experimental set-up I, other experiments were run with $^1$D flow 0.1 ml/min, $^2$D flow was increased to 20 ml/min and modulation period 7 s, while modulation injection time was varied. Calculations are reported in Table SIV in the Supporting material for these conditions. As it can be seen from the value of $F_{(L)}$, optimal conditions required injection time of 0.12 s. The corresponding chromatograms, as expected, demonstrate very nice peaks with minimal tailing and fronting (Figure S5A and S5B), illustrating that the score function is still relevant.

A second set of analysis was performed with set-up II consisting in a DB-1 (20 m, 0.1 mm ID, 0.4 µm) and a BPX-50 (5 m, 0.25 mm ID, 0.25 µm) column. This is a more appropriate set-up for flow modulation as a longer $^{2}$D column decreases the change of the average velocity in the modulation channel from the beginning to the end of the thermally programmed GC run. In this configuration, $^{1}$D flow of 0.1 ml/min, $^{2}$D flow of 8 ml/min and modulation period of 5.5 s were applied with different injection times from 0.14 to 0.27 s. Additional experiments were run with He as a carrier gas, involving $^{1}$D flow 0.15 ml/min, $^{2}$D flow 27 ml/min and modulation period 8 s or $^{1}$D flow 0.15 ml/min, $^{2}$D flow 25 ml/min and modulation period 6 s. For all these conditions, the score function was calculated with various modulation injection times. All the obtained results are reported in Table SV to SVII in the Supporting material. For each of the conditions, minimum values of the score function could be identified and experimentally led to satisfactory peak shapes all along the entire GC run, contrary to high values of the score function. In this configuration, carrier gas was also changed from $H_2$ to He in order to check the prediction performance of the calculations for a different carrier gas. Same conclusions on the ability of the score function to predict adequate conditions were drawn.

Numerous analysis were also performed with set-up III: Rxi-1ms (30 m, 0.25 mm ID, 0.25 µm) and ZB-35HT (5 m, 0.25 mm ID, 0.18 µm) columns. This set up is often employed with forward fill/flush flow modulation, however it can be problematic owing to high $^{1}$D flows which are necessarily employed due to higher ID of $^{1}$D column, as this causes flush/fill ratio to be low which might compromise the efficiency of modulation. The score function gave optimal results for $^{1}$D flow 0.3 ml/min, $^{2}$D flow 18 ml/min and modulation period 4 s, with injection times between 0.1 and 0.16 s (Table SVIII in Supplementary material). Once again, minimum value of the function corresponded to adequate conditions leading to appropriate peak shapes.

On this set-up, some predicted optimal conditions were applied ($^{1}$D flow 0.3 ml/min, $^{2}$D flow 27 ml/min, modulation period 6 s and injection time of 0.1 s) and chromatograms were acquired at different oven ramps: 1.5, 2.5 and 3 °C/min. Modulated peaks over chromatograms were compared (Figure S14 in Supplementary material). No significant difference in the shapes of the peaks was noticed testifying that the score function can be safely employed for usual GC×GC oven ramps.

In the end, analysis were also performed with ZB-5HT (15 m, 0.1 mm ID, 0.1 µm) × ZB-35HT (5 m, 0.25 mm ID, 0.18 µm) columns set IV, which is a set-up adapted for the analysis of samples containing very high boiling point analytes. A test mix with a wider range of carbon number, from n-$C_{14}$ to n-$C_{44}$ was employed. Since $^{1}$D column in this case has a thin stationary phase, it is very easily overloaded with sample and $^{1}$D peaks can easily demonstrate significant fronting.

To check the performance of the score function in those high temperature conditions (up to 350°C), $^{1}$D flow was set to 0.12 ml/min, $^{2}$D flow to 12 ml/min and modulation period to 7 s. The best injection time according to the calculations was estimated to be around 0.16-0.18 s. Once again, the best peak shapes were obtained for the lowest values of the score function. Deviating from these optimal values lead to visible alteration of peak shape (see Supplementary material).

In all cases minimising the value of the score function gave good prediction for the most optimal conditions. Overall, it was observed that values of function less then ~10 gave still satisfactory peak shapes, while when the value of the score function was higher than 20 serious deterioration of the peak shapes was observed. High flush to fill distance ratio further improves modulation performance, hence even if $F_{(L)}$ is slightly higher and Flush/Fill ratio is higher than 5, good peak shapes can still be obtained. In general, low values for the flush to fill distance ratio (<1.5) should be avoided. Precisely for this reason, high modulation periods can be detrimental for quantification performance in forward fill/flush modulation as they result in high fill distances, hence possibly lower Flush/Fill ratios. However, high flush to fill distance ratio often implies low modulation periods and low sampling times that in turn require proper choice of the separating conditions to avoid oversampling and/or wrap-around.

### 3.4.4. Literature operating conditions evaluation

To evaluate the interest of this score function for the scientific community, it was decided to calculate the values of the $F_{(L)}$ score function for the published GC×GC conditions given in Table I (see last column of the table). While our intention is not to judge the quality of our peers' work, it is interesting to note that obtained values are quite different from one study to another, from satisfactory values inferior to 10 to very large values above 100. These high values will indicate peak shape aggravation and quantification issues.

### 3.5. Flow modulation calculator

In order to provide to a user with a generic tool that would help in choosing optimal modulation conditions, a dedicated calculator was designed. Code written in Python 3.7 allows to perform following set of steps (Figure 6). The code and the executable are available in Supporting Material, in this way any user can check its modulator performance and calibrate the tool possibly if any differences in modulator channel length or diameter are observed.

Interface is easy to use and is illustrated in Figure 7. Mandatory parameters are dimensions of $^{1}$D and $^{2}$D columns, analysis initial and final temperatures. $^{1}$D flow is also chosen to be one of the mandatory variables as well as the nature of the carrier gas. A first possibility of use for this calculator is to use it as a simple tool to check if operating conditions are satisfactory. Adequate conditions result in green colors ($F_{(L)}$ values ≤ 10 or Flush/Fill distance ratio ≥ 2.5), non-ideal conditions in orange ($10 < F_{(L)} < 20$ or Flush/Fill distance ratio < 2.5) and unsatisfactory conditions in red ($F_{(L)} \geq 20$ or Flush/Fill distance ratio ≤1) – see Figure 7. A second option is to use the calculator as an optimization tool by leaving some operating parameters as floating ones. $^{2}$D flow, modulation period and injection time are concerned in this case. If these are not defined and left to the '0' default value, the calculator estimates the best combination of parameters based on the minimum $F_{(L)}$ value and reports back suggested values of the three mentioned parameters. Since however many combinations of parameters can result in a similar $F_{(L)}$ value, better performance of the calculator is obtained if only one or two parameters are estimated.

## 4. Conclusion

In this work, the behaviour of the forward/fill flush flow modulator was investigated with different column sets and a large panel of operating parameters. Modulation is generally a very sensitive process which can be affected by many parameters and that requires time-consuming lab experiments. However, based on hundreds of experiments, it was demonstrated that modulation performance all along the GC run can be predicted with a good success based on the physical rules dictating its performance. A score function which contains embedded major descriptors of the modulation process (travelled distances of the GC effluent in the modulation channel) and which awards penalty points if these descriptors deviate from the optimal values was designed. Calculating score function value for many conditions and choosing the one that leads to minimum score function value is demonstrated to be a good way to choose most optimal modulation conditions. In the end, this function was embedded in a calculator which allows the users to quickly check the expected performance of their modulation process and/or choose appropriate conditions.

**Useful links**

Flow modulation calculator "exe" file can be obtained here.

**List of figures:**

Figure 1 A) n-$C_{10}$ measured modulated peak-traced in black. Calculated signal with fitted gaussians-traced in red. Residuals between the two signals-dashed blue line. B) Zoom on a single peak area.

Figure 2 Resulting PLS loadings plot for dependent (in blue) and independent variables (in red).

Figure 3 Plot of the function $F_{(L)}$ for all the combinations of the chosen set of modulation parameters. Insert: $F_{(L)}$ values up to ca. 40.

Figure 4 A) n-$C_{10}$ modulated peaks for 0.14, 0.15, 0.16 s injection times; B) n-$C_{28}$ modulated peaks for 0.14, 0.15, 0.16 s injection times; C) n-$C_{10}$ modulated peaks for 0.12, 0.15, 0.18 s injection times; D) n-$C_{28}$ modulated peak for 0.12, 0.15, 0.18 s injection times; E) n-$C_{10}$ modulated peak for 0.23 s injection time; F) n-$C_{28}$ modulated peak for 0.23 s injection time. $^1$D flow 0.15 ml/min, $^2$D flow 11.7 ml/min, modulation period 4.5 s.

Figure 5 A) n-$C_{10}$ modulated peaks for $^2$D flow 6 ml/min; B) n-$C_{28}$ modulated peaks for $^2$D flow 6 ml/min; C) n-$C_{10}$ modulated peaks for $^2$D flow 10 ml/min; D) n-$C_{28}$ modulated peaks for $^2$D flow 10 ml/min, E) n-$C_{10}$ modulated peaks for $^2$D flow 14 ml/min; F) n-$C_{28}$ modulated peaks for $^2$D flow 14 ml/min. $^1$D flow 0.1 ml/min, modulation period 5.5 s, injection time 0.18 s.

Figure 6 Flowchart describing the steps of the program leading to the estimation of $F_{(L)}$.

Figure 7 Flow modulation calculator. Green colour indicates acceptable values ($F_{(L)} \leq 10$), orange- a warning is issued ($F_{(L)} > 10$ and $F_{(L)} < 20$), while red indicates possible serious modulation performance issues ($F_{(L)} \geq 20$).

**List of tables:**

**Table I Overview of the operating conditions of the GC×GC separation involving forward fill/flush differential flow modulator found in literature.***

| Sample type | Column set | Carrier gas | Flows | Modulation period and injection time | Oven rate | Ref. | Calculated $F_{(L)}$ and Flush/Fill** |
|---|---|---|---|---|---|---|---|
| Bacterial fatty acids | HP-5MS (30 m, 0.25 mm, 0.25 µm)<br>BPX-70 (4 m, 0.25 mm, 0.25 µm) | $H_2$ | 0.6 mL/min<br>25 mL/min | 2.0 s<br>0.1 s injection | 2 °C/min | [14] 2010 | Dual detection-not calculated |
| Fatty acids | DB-1MS (10 m, 0.1 mm, 0.1 µm)<br>SLB-IL 82, SLB-IL 100 and HP-88 (4 m, 0.25 mm, 0.2 µm) | $H_2$ | 0.3 mL/min<br>24 mL/min | 2 s<br>0.1 s injection | 10 °C/min and 3 °C/min | [15] 2011 | $F_{(L)}$=18.93<br>Fill/Flush=4.2<br>$L^{Tin}$=11.9 $L^{Tfin}$=13.1 |
| Light cycle oil | DB-5 (10 m, 0.1 mm, 0.4 µm)<br>BPX-50 (10 m, 0.25 mm, 0.1 µm) | $H_2$ | 0.2 mL/min<br>22 mL/min | 11 s<br>0.2 s injection | 1.9 °C/min | [6] 2011 | $F_{(L)}$==32.13<br>Fill/Flush=2.04<br>$L^{Tin}$=17.7 $L^{Tfin}$=20.4 |
| Gasoline, reformate and fluid catalytic cracking samples | SLB-IL 111 (30 m, 0.25 mm, 0.2 µm)<br>HP-5MS (5 m, 0.25 mm, 0.25 µm) | He | 0.6 mL/min<br>23 mL/min | 6.0 s<br>0.52 s injection | 3 °C/min | [16] 2013 | Dual detection-not calculated |
| Fatty Acid | DB-1MS (20 m, 0.1 mm, 0.1 µm)<br>HP-INNOWax (5 m, 0.25 mm, 0.15 µm) | $H_2$ | 0.3 mL/min<br>24 mL/min | 2 s<br>0.25 s injection | 5 °C/min | [17] 2014 | $F_{(L)}$=111.62<br>Fill/Flush=11.42<br>$L^{Tin}$=22.7 $L^{Tfin}$=26.3 |
| VGO samples | ZB1-HT (15 m, 0.1 mm, 0.1 µm)<br>ZB35-HT (5 m, 0.25 mm, 0.1 µm) | $H_2$ | 0.15 mL/min<br>28 mL/min | 8.965 s<br>0.2 s injection | 2 °C/min | [18] 2014 | $F_{(L)}$=115.72<br>Fill/Flush=4.25<br>$L^{Tin}$=22.6 $L^{Tfin}$=26.9 |
| Petroleum reformate product and essential oil | DB-5MS (30 m, 0.25 mm, 0.25 µm)<br>Rt-βDEXse (30 m, 0.25 mm, 0.25 µm)<br><br>HP-INNOWax (5 m, 0.25 mm, 0.15 µm) | He | 0.7 mL/min<br>23 mL/min | 6.0 s<br>0.52 s injection | 2 °C/min and 3 °C/min | [19–21] 2013-2016 | Dual detection-not calculated |
| VGO samples | DB5-HT (10 m, 0.1 mm, 0.1 µm)<br>ZB35-HT Inferno (5 m, 0.25 mm, 0.1 µm) | $H_2$ | 0.14 mL/min<br>24 mL/min | 8 s<br>0.2 s injection | 1.9 °C/min | [22] 2015 | $F_{(L)}$=91.31<br>Fill/Flush=4.39<br>$L^{Tin}$=20.5 $L^{Tfin}$=24.9 |
| Fatty Acid | SP-2560 (75 m, 0.18 mm, 0.14 µm)<br>19091-L431 (3.5 m, 0.25 mm, 0.14 µm) | $H_2$ | 0.2 mL/min<br>22 mL/min | 2.9 s<br>0.13 s injection | 2 °C/min | [23–25] 2015-2017 | $F_{(L)}$=2.1<br>Fill/Flush=5.15<br>$L^{Tin}$=11.9 $L^{Tfin}$=13.1 |
| VOCs in breath gas | Rxi-5Sil MS (30 m, 0.25 mm, 0.2 µm)<br>DB-WAX (4 m, 0.25 mm, 0.25 µm) | He | 0.6 mL/min<br>23 mL/min | 3.0 s<br>0.201 s injection | 3 °C/min and 5 °C/min | [26] 2019 | Dual detection-not calculated |
| Diesel oils | Rxi-5MS (30 m, 0.25 mm, 0.25 µm)<br>HP-INNOWax (5 m, 0.25 mm, 0.1 µm) | $H_2$ | 0.8 mL/min<br>25 mL/min | 1.52 s<br>0.12 s injection | 8 °C/min | [27] 2020 | $F_{(L)}$=3.47<br>Fill/Flush=2.68<br>$L^{Tin}$=14.1 $L^{Tfin}$=16.3 |

* **HP-5MS** - 5%-phenyl-methylpolysiloxane; **BPX-70** - 70%-cyanopropyl polysilphenylene-siloxane; **DB-1MS** - 100% dimethylpolysiloxane; **SLB-IL 82** - 1,12-Di(2,3-dimethylimidazolium)dodecane bis(trifluoromethanesulfonyl)imide; **SLB-IL 100** - 1,9-Di(3-vinylimidazolium)nonane bis(trifluoromethanesulfonyl)imide; **HP-88** - 88% Cyanopropy)aryl-polysiloxane; **DB-5** - 5% Diphenyl / 95% Dimethylpolysiloxan; **BPX-50** - 50% Phenyl Polysilphenylene-siloxane; **SLB-IL 111** - 1,5-Di(2,3-dimethylimidazolium)pentane bis(trifluoromethanesulfonyl)imide; **HP-5MS** - 5%-phenyl-methylpolysiloxane; **HP-INNOWax** - polyethylene glycol; **ZB1 HT** - 100 % dimethylpolysiloxane; **ZB35 HT** - 35 %-phenyl65 %-dimethylpolysiloxane; **Rt-βDEXse** - 2,3-di-O-ethyl-6-O-tert-butyl dimethylsilyl beta cyclodextrin added into 14% cyanopropylphenyl/86% dimethyl polysiloxane; **SP-2560** - poly(biscyanopropyl siloxane); **19091-L431** – (50%-phenyl)-methylpolysiloxane; **Rxi-5Sil MS** - Crossbond 1,4-bis(dimethylsiloxy)phenylene dimethyl polysiloxane; **DB-WAX** - Polyethylene glycol.

** Initial and final oven temperatures are used for the estimation of $F_{(L)}$.

**Table II Overview of all modulation conditions presented in this work.**

| Column set | 1D Flow (ml/min) | 2D Flow (ml/min) | Modulation period (s) | Injection time (s) | Carrier gas |
|---|---|---|---|---|---|
| **I:** DB-1 (20 m, 0.1 mm, 0.4 µm) × BPX-50 (3.2 m, 0.25 mm, 0.25 µm) | 0.15 | 11.7 | 4.5 | 0.12-0.23 | $H_2$ |
| | 0.1 | 6-18 | 5.5 | 0.18 | $H_2$ |
| | 0.1 | 20 | 7 | 0.05-0.18 | $H_2$ |
| **II:** DB-1 (20 m, 0.1 mm, 0.4 µm) × BPX-50 (5 m, 0.25 mm, 0.25 µm) | 0.1 | 8 | 5.5 | 0.14-27 | $H_2$ |
| | 0.15 | 27 | 8 | 0.14-0.25 | He |
| | 0.15 | 25 | 6 | 0.3 | He |
| **III:** Rxi-1ms (30 m, 0.25 mm, 0.25 µm) × ZB-35HT (5 m, 0.25 mm, 0.18 µm) | 0.3 | 18 | 4 | 0.1-0.16 | $H_2$ |
| | 0.3 | 27 | 6 | 0.1 | $H_2$ |
| **IV:** ZB-5HT (15 m, 0.1 mm, 0.1 µm) × ZB-35HT (5 m, 0.25 mm, 0.18 µm) | 0.12 | 12 | 7 | 0.14-0.26 | $H_2$ |

**Table III Calculated values of traversed fill and flush distances and associated $F_{(L)}$ value for various modulation injection times, 1D flow 0.15 ml/min, 2D flow 11.7 ml/min, modulation period 4.5 s.**

| Mod injection time (s) | Flush/Fill | Fill distance (cm) | Flush distance(cm) | Fill +Flush (cm) | Fill distance (cm) | Flush distance(cm) | Fill +Flush (cm) | $F_{(L)}$ |
|---|---|---|---|---|---|---|---|---|
| | | 100°C | | | 300°C | | | |
| 0.12 | 2.13 | 3.73 | 7.96 | 11.69 | 4.43 | 9.45 | 13.89 | 16.07 |
| 0.14 | 2.50 | 3.71 | 9.28 | 12.99 | 4.41 | 11.03 | 15.44 | 4.14 |
| 0.15 | 2.68 | 3.70 | 9.94 | 13.64 | 4.40 | 11.82 | 16.22 | 4.30 |
| 0.16 | 2.87 | 3.69 | 10.61 | 14.30 | 4.39 | 12.60 | 17.00 | 4.45 |
| 0.17 | 3.06 | 3.69 | 11.27 | 14.96 | 4.38 | 13.39 | 17.77 | 7.38 |
| 0.18 | 3.24 | 3.68 | 11.93 | 15.61 | 4.37 | 14.18 | 18.55 | 11.37 |
| 0.19 | 3.43 | 3.67 | 12.59 | 16.26 | 4.36 | 14.97 | 19.33 | 19.37 |
| 0.23 | 4.19 | 3.63 | 15.25 | 18.88 | 4.32 | 18.12 | 22.44 | 66.35 |

**Table IV Calculated values of traversed fill and flush distances and associated $F_{(L)}$ value for various 2D flows, 1D flow 0.1 ml/min, modulation period 5.5 s, injection time 0.18 s.**

| 2D flow (ml/min) | Flush/Fill | Fill distance (cm) | Flush distance(cm) | Fill +Flush (cm) | Fill distance (cm) | Flush distance(cm) | Fill +Flush (cm) | $F_{(L)}$ |
|---|---|---|---|---|---|---|---|---|
| | | 100°C | | | 300°C | | | |
| 6.00 | 2.03 | 3.62 | 7.33 | 10.95 | 4.55 | 9.22 | 13.77 | 31.1 |
| 8.00 | 2.70 | 3.37 | 9.10 | 12.82 | 4.13 | 11.16 | 15.29 | 4.77 |
| 10.00 | 3.38 | 3.17 | 10.69 | 13.86 | 3.81 | 12.87 | 16.68 | 4.44 |
| 12.00 | 4.05 | 2.99 | 12.14 | 15.13 | 3.55 | 14.40 | 17.95 | 7.61 |
| 14.00 | 4.73 | 2.85 | 13.47 | 16.32 | 3.34 | 15.81 | 19.15 | 18.53 |
| 16.00 | 5.40 | 2.72 | 14.72 | 17.44 | 3.17 | 17.11 | 20.27 | 27.94 |
| 18.00 | 6.08 | 2.61 | 15.88 | 18.49 | 3.01 | 18.32 | 21.34 | 56.46 |

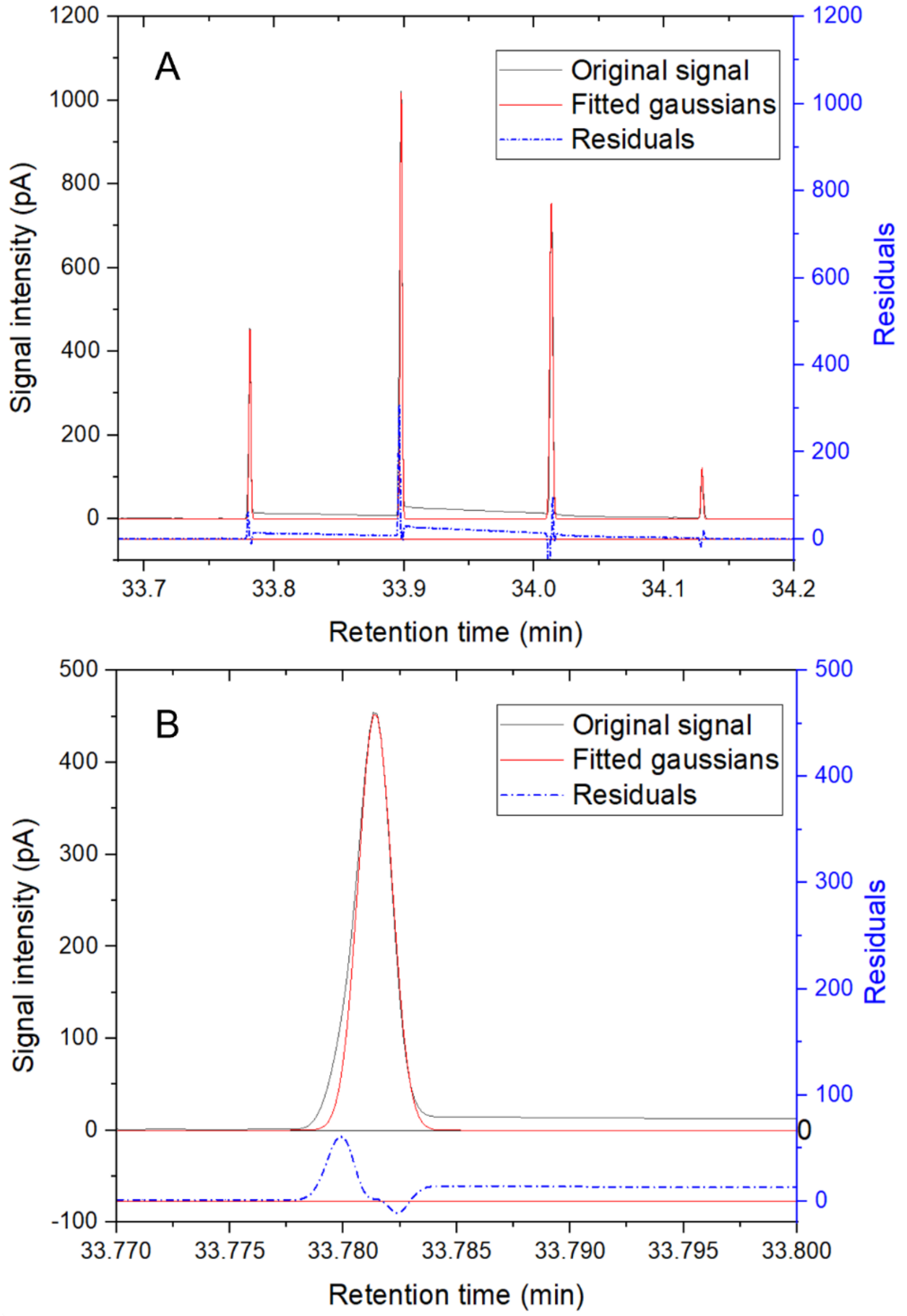

Figure 1 A) n-$C_{10}$ measured modulated peak-traced in black. Calculated signal with fitted gaussians-traced in red. Residuals between the two signals-dashed blue line. B) Zoom on a single peak area.

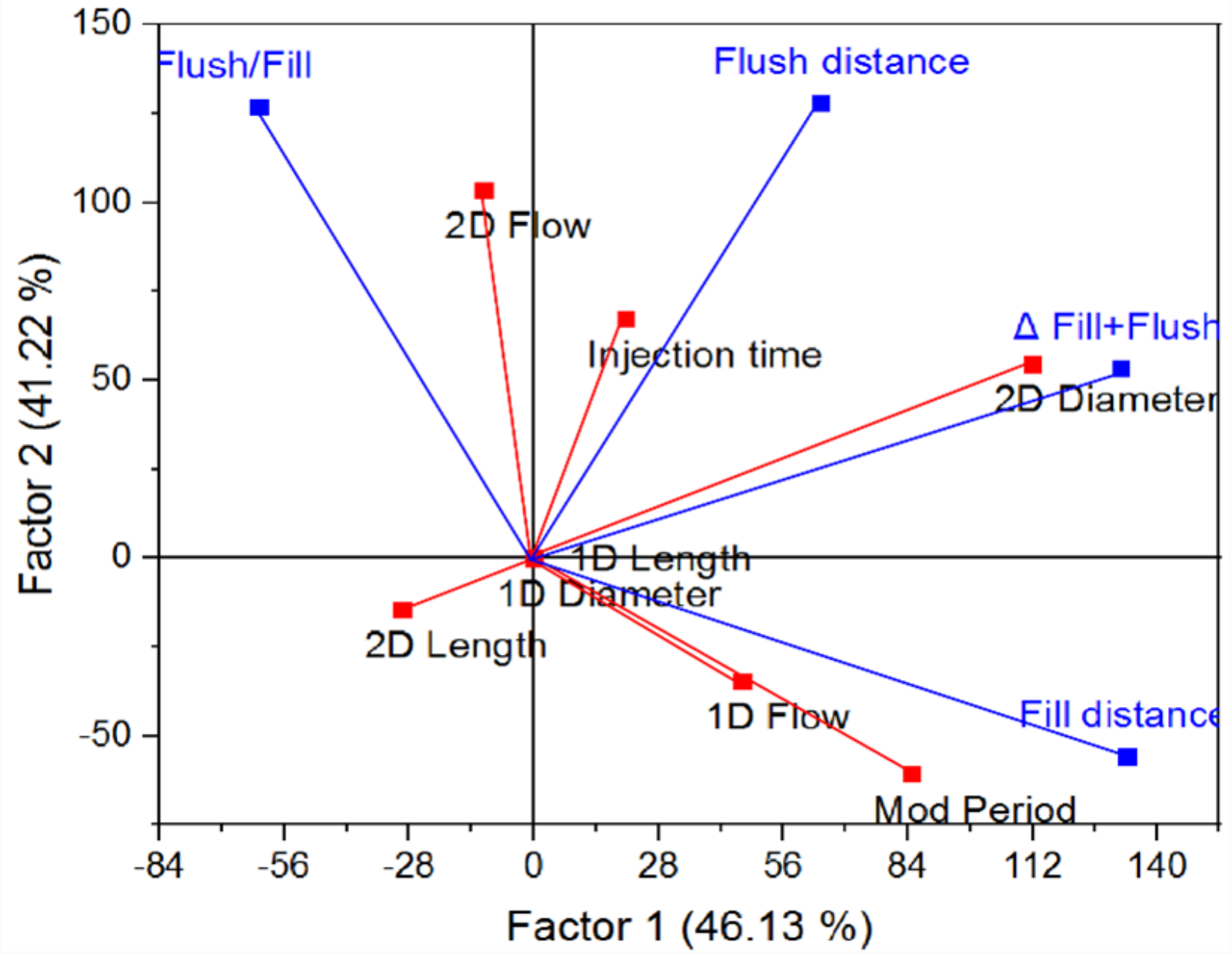

Figure 2 Resulting PLS loadings plot for dependent (in blue) and independent variables (in red).

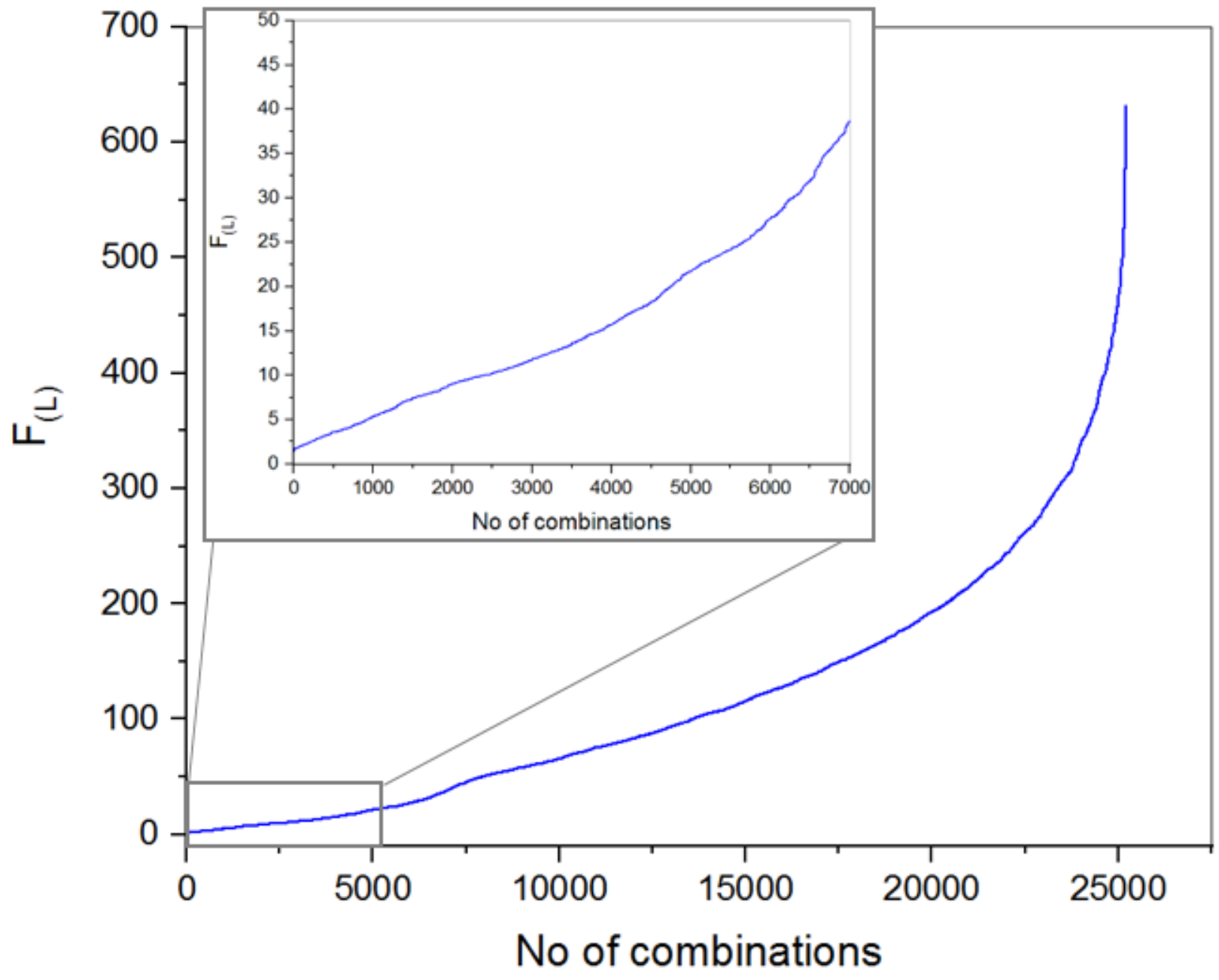

Figure 3 Plot of the function $F_{(L)}$ for all the combinations of the chosen set of modulation parameters. Insert: $F_{(L)}$ values up to ca. 40.

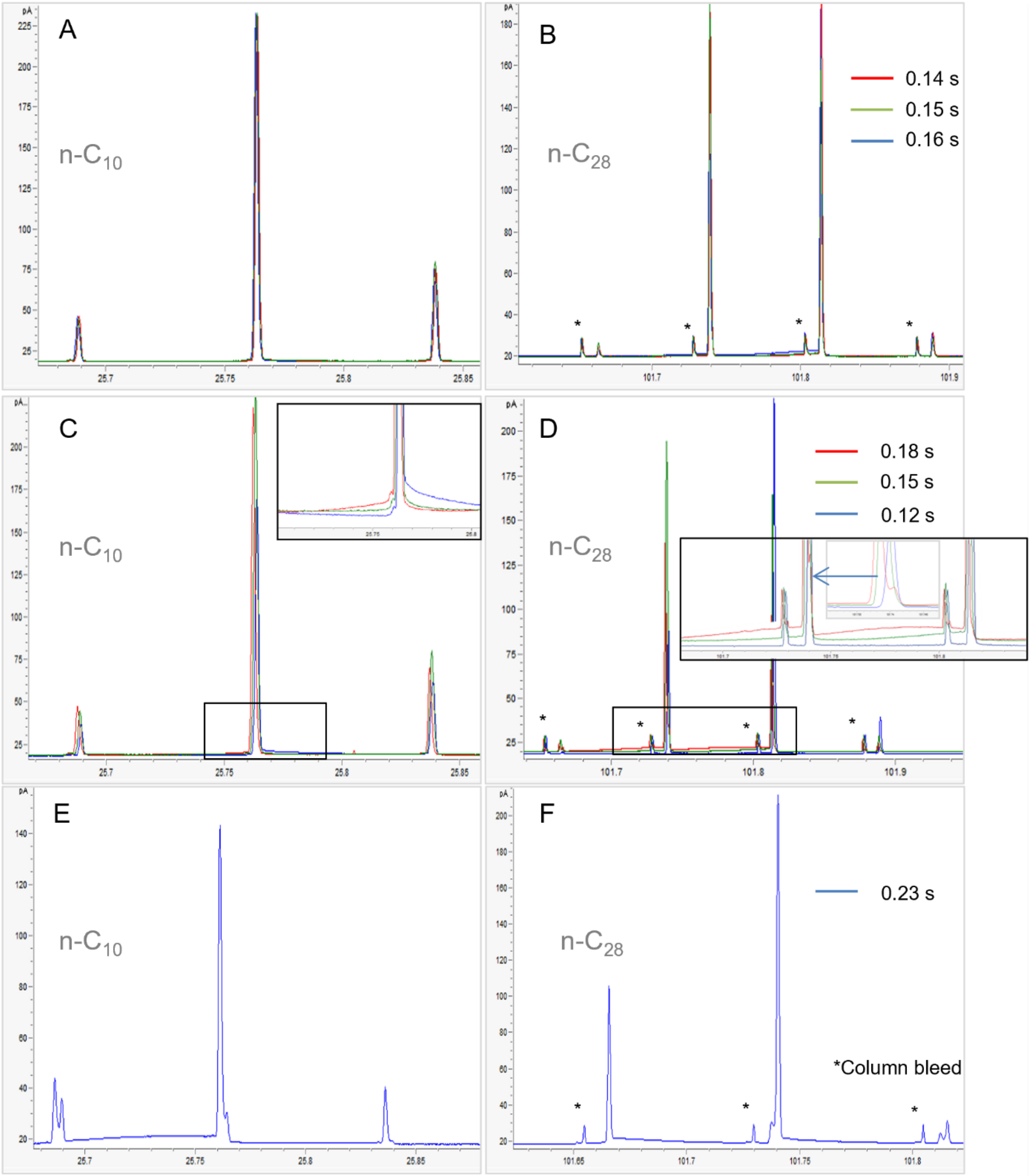

Figure 4 A) n-$C_{10}$ modulated peaks for 0.14, 0.15, 0.16 s injection times; B) n-$C_{28}$ modulated peaks for 0.14, 0.15, 0.16 s injection times; C) n-$C_{10}$ modulated peaks for 0.12, 0.15, 0.18 s injection times; D) n-$C_{28}$ modulated peak for 0.12, 0.15, 0.18 s injection times; E) n-$C_{10}$ modulated peak for 0.23 s injection time; F) n-$C_{28}$ modulated peak for 0.23 s injection time. $^{1}$D flow 0.15 ml/min, $^{2}$D flow 11.7 ml/min, modulation period 4.5 s.

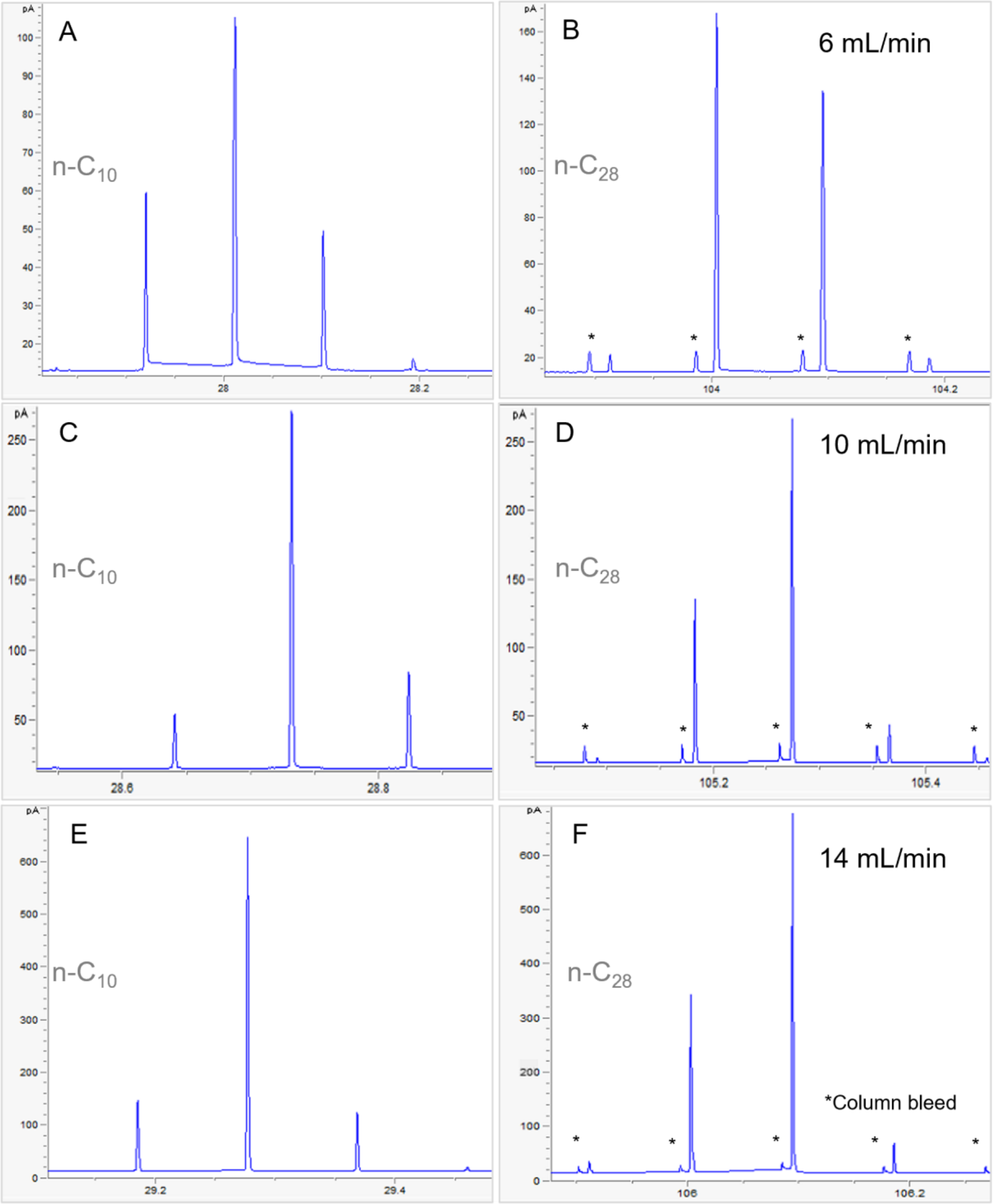

Figure 5 A) n-$C_{10}$ modulated peaks for $^2$D flow 6 ml/min; B) n-$C_{28}$ modulated peaks for $^2$D flow 6 ml/min; C) n-$C_{10}$ modulated peaks for $^2$D flow 10 ml/min; D) n-$C_{28}$ modulated peaks for $^2$D flow 10 ml/min, E) n-$C_{10}$ modulated peaks for $^2$D flow 14 ml/min; F) n-$C_{28}$ modulated peaks for $^2$D flow 14 ml/min. $^1$D flow 0.1 ml/min, modulation period 5.5 s, injection time 0.18 s.

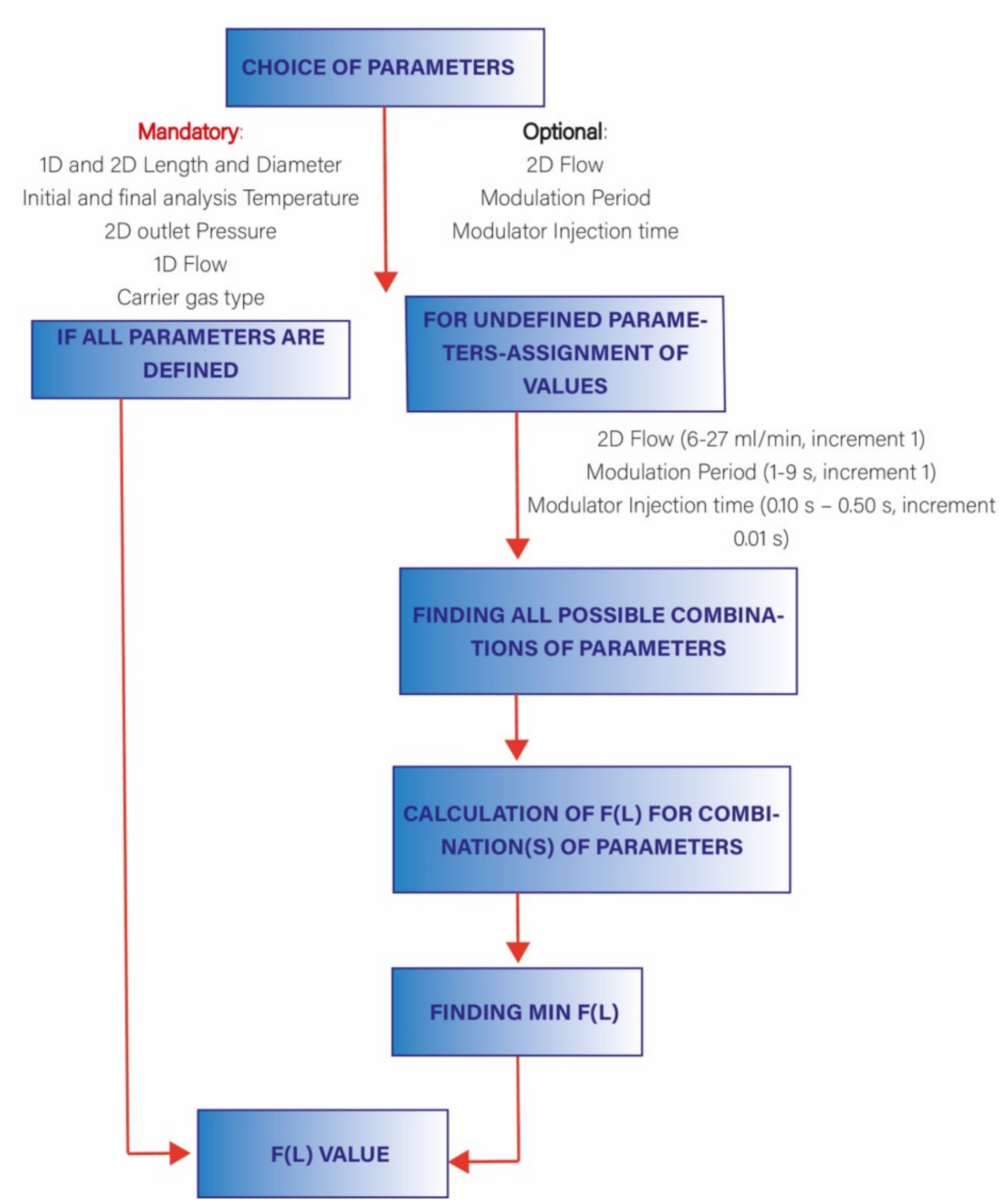

Figure 6 Flowchart describing the steps of the program leading to the estimation of $F_{(L)}$.

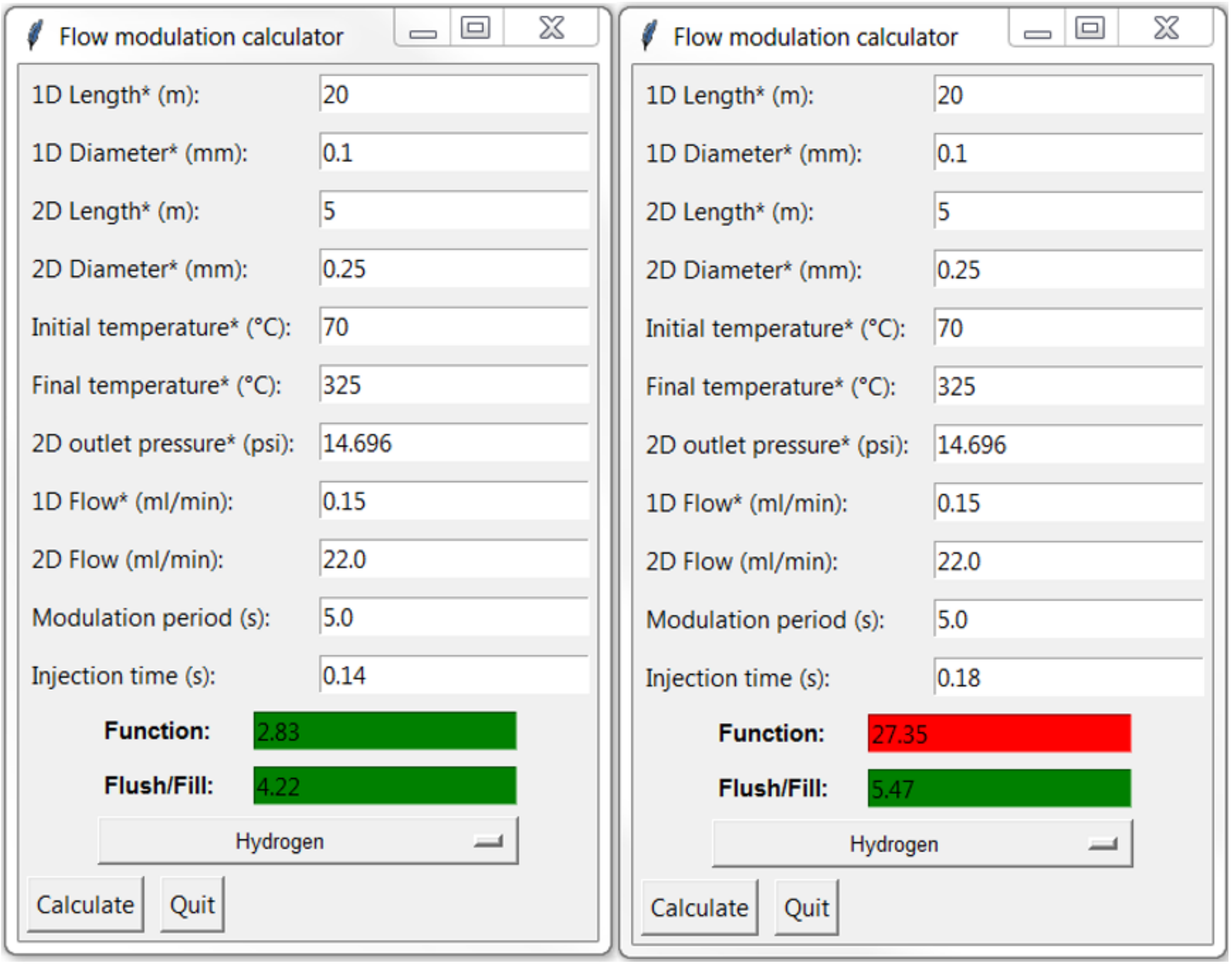

Figure 7 Flow modulation calculator. Green colour indicates acceptable values ($F_{(L)} \leq 10$), orange- a warning is issued ($F_{(L)} > 10$ and $F_{(L)} < 20$), while red indicates possible serious modulation performance issues ($F_{(L)} \geq 20$).

# Supplementary material

## Score function for the optimization of the performance of forward fill/ flush differential flow modulation for comprehensive two-dimensional gas chromatography

Aleksandra Lelevic[a,b,*], Christophe Geantet[b], Chantal Lorentz[b], Maxime Moreaud[a], Vincent Souchon[a]

a. IFP Energies nouvelles, Rond-point de l'échangeur de Solaize BP 3 69360 Solaize France

b. Univ Lyon, Université Claude Bernard Lyon 1, CNRS, IRCELYON, F-69626, Villeurbanne, France

* Author for correspondence: aleksandra.lelevic@ifpen.fr

**Table SI Composition of the n-paraffin Test mixture (n-C10 to n-C28).**

| Compound | m/m% |
|---|---|
| n-C10 | 0.12 |
| n-C12 | 0.11 |
| n-C16 | 0.12 |
| n-C18 | 0.13 |
| n-C20 | 0.10 |
| n-C22 | 0.10 |
| n-C24 | 0.10 |
| n-C26 | 0.10 |
| n-C28 | 0.11 |

**Table SII Composition of the n-paraffin Test mixture (n-C10 to n-C44).**

| Compound | m/m% |
|---|---|
| n-C10 | 10.39 |
| n-C12 | 8.24 |
| n-C15 | 13.33 |
| n-C16 | 12.66 |
| n-C18 | 4.05 |
| n-C20 | 3.49 |
| n-C22 | 14.22 |
| n-C24 | 5.49 |
| n-C26 | 5.46 |
| n-C28 | 4.79 |
| n-C30 | 4.97 |
| n-C32 | 3.49 |
| n-C36 | 4.53 |
| n-C38 | 3.19 |
| n-C44 | 1.71 |

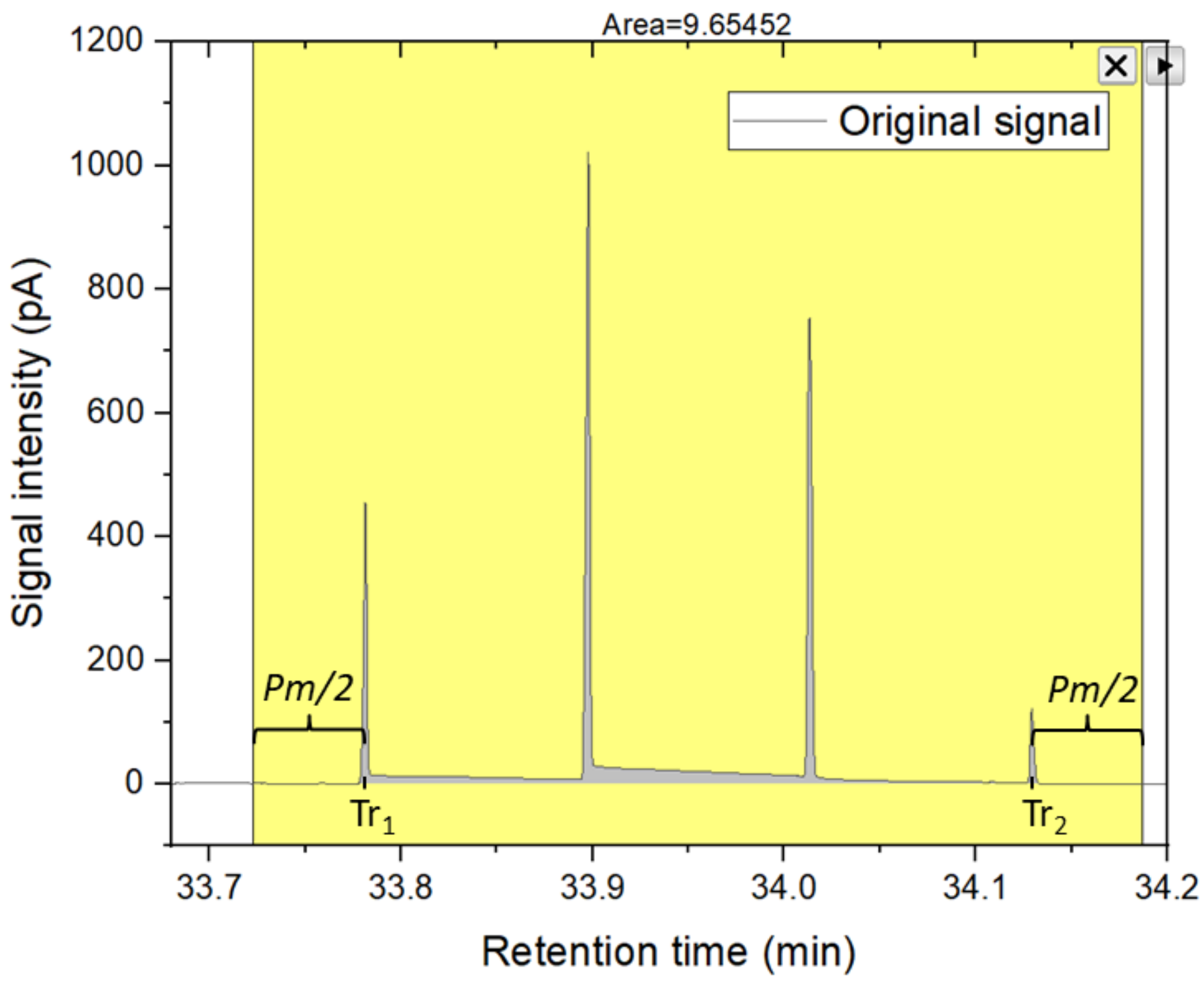

**Figure S1 Modulated peak signal normalization is obtained by dividing the signal at every point by total modulated peak area integrated as illustrated in yellow area.**

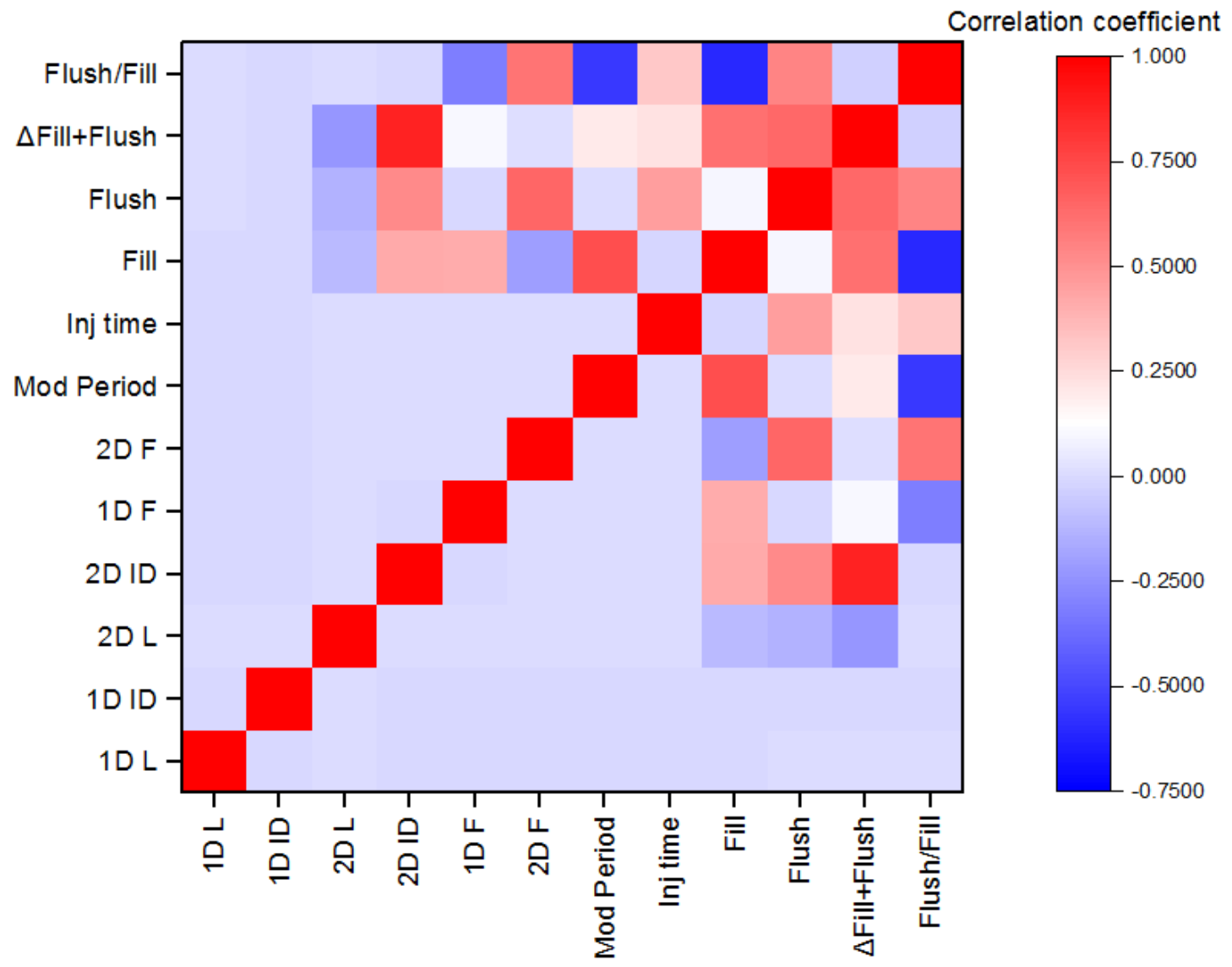

**Figure S2 Pearson correlation coefficient for investigated modulation variables and modulation descriptors.**

**Table SIII Overview of all the modulation conditions reported in the validation Section.**

| Column set | 1D Flow (ml/min) | 2D Flow (ml/min) | Modulation period (s) | Injection time (s) | Carrier gas |
|---|---|---|---|---|---|
| **I:** DB-1 (20 m, 0.1 mm , 0.4 µm) × BPX-50 (3.2 m, 0.25 mm , 0.25 µm) | 0.15 | 11.7 | 4.5 | 0.12-0.23 | $H_2$ |
| | 0.1 | 6-18 | 5.5 | 0.18 | $H_2$ |
| | 0.1 | 20 | 7 | 0.05-0.18 | $H_2$ |
| **II:** DB-1 (20 m, 0.1 mm , 0.4 µm) × BPX-50 (5 m, 0.25 mm , 0.25 µm) | 0.1 | 8 | 5.5 | 0.14-27 | $H_2$ |
| | 0.15 | 27 | 8 | 0.14-0.25 | He |
| | 0.15 | 25 | 6 | 0.3 | He |
| **III:** Rxi-1ms (30 m, 0.25 mm , 0.25 µm) × ZB-35HT (5 m, 0.25 mm , 0.18 µm) | 0.3 | 18 | 4 | 0.1-0.16 | $H_2$ |
| | 0.3 | 27 | 6 | 0.1 | $H_2$ |
| **IV:** ZB-5HT (15 m, 0.1 mm , 0.1 µm) × ZB-35HT (5 m, 0.25 mm , 0.18 µm) | 0.12 | 12 | 7 | 0.14-0.26 | $H_2$ |

*Set-up **I**: DB-1 (20 m, 0.1 mm ID, 0.4 µm) × BPX-50 (3.2 m, 0.25 mm ID, 0.25 µm)*
*Carrier gas: $H_2$*

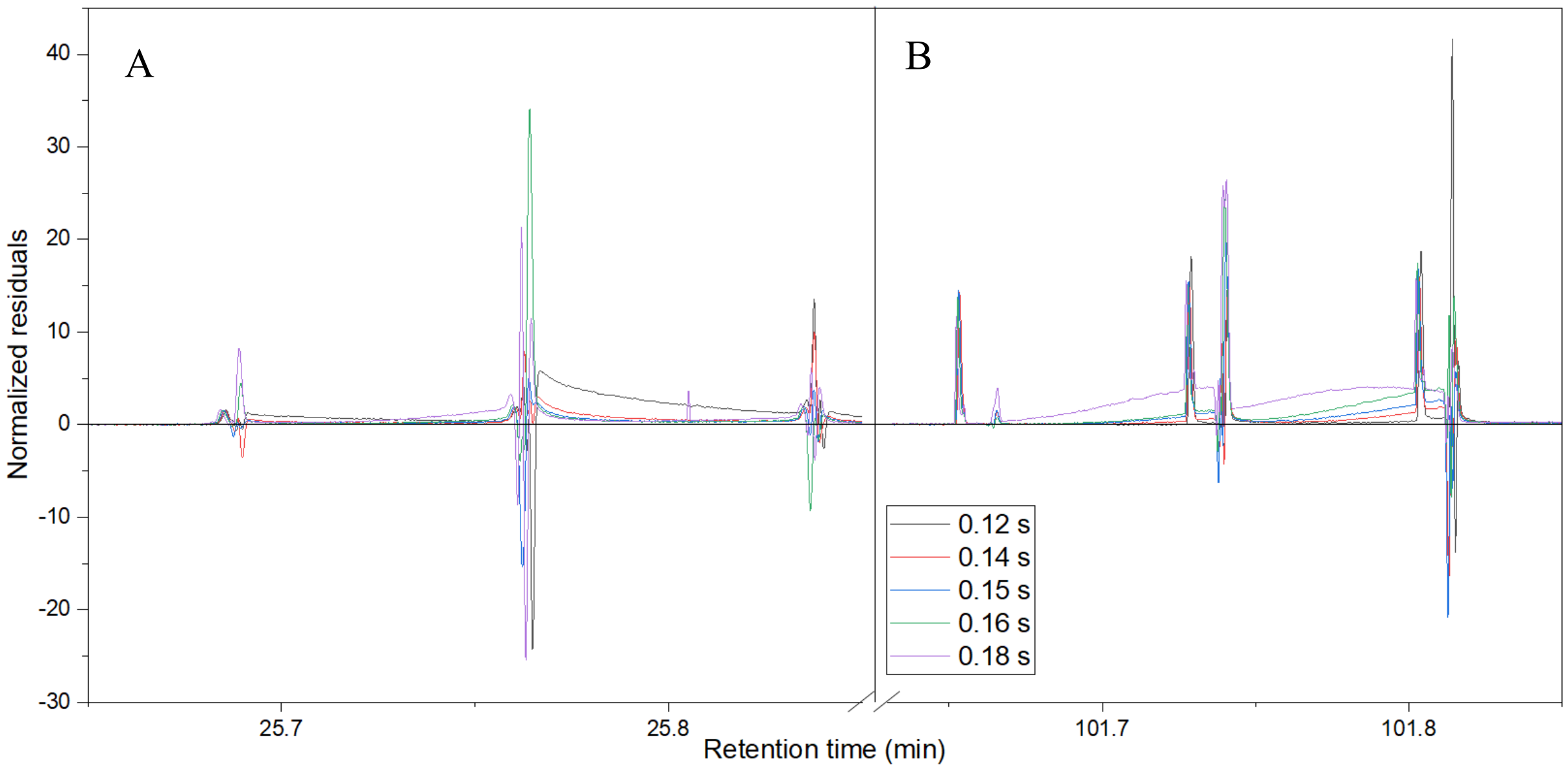

**Figure S3 Normalized residuals for A) n-C10 modulated peak; B) n-C28 modulated peak. 1D flow 0.15 ml/min, 2D flow 11.7 ml/min, modulation period 4.5 s, injection times 0.12-0.18 s.**

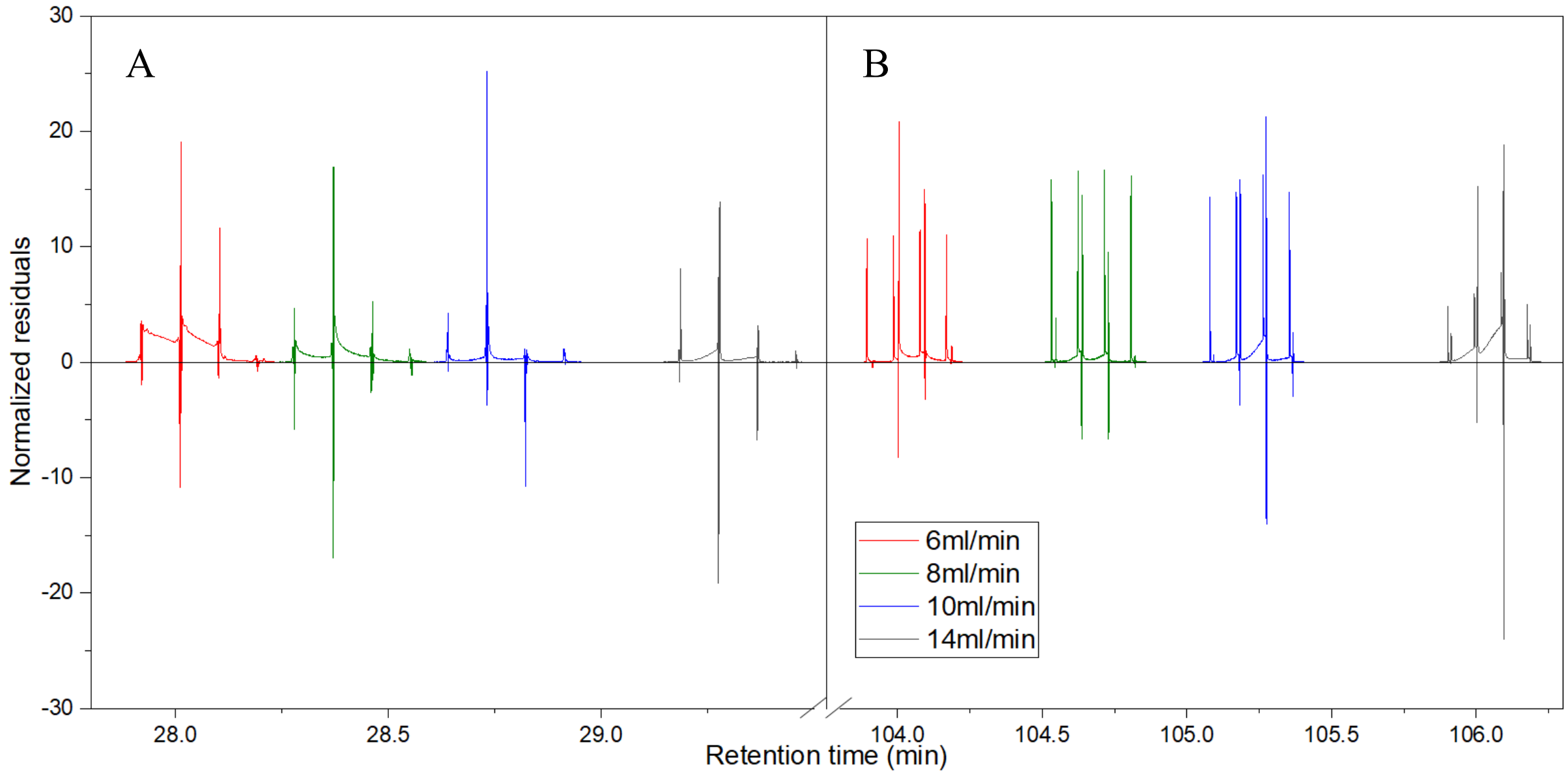

**Figure S4 Normalized residuals for A) n-C10 modulated peak; B) n-C28 modulated peak. 1D flow 0.1 ml/min, modulation period 5.5 s, injection time 0.18 s, 2D Flows 6-14 ml/min.**

| Mod injection time (s) | Flush/Fill | Fill distance (cm) | Flush distance(cm) | Fill +Flush (cm) | Fill distance (cm) | Flush distance(cm) | Fill +Flush (cm) | $F_{(L)}$ |
|---|---|---|---|---|---|---|---|---|
| | | 100°C | | | 300°C | | | |
| 0.05 | 1.44 | 3.28 | 4.72 | 8.00 | 3.76 | 5.41 | 9.17 | 76.75 |
| 0.10 | 2.89 | 3.26 | 9.43 | 12.69 | 3.74 | 10.82 | 14.55 | 4.56 |
| 0.12 | 3.48 | 3.25 | 11.32 | 14.57 | 3.73 | 12.98 | 16.71 | 3.03 |
| 0.14 | 4.07 | 3.24 | 13.20 | 16.44 | 3.72 | 15.14 | 18.86 | 18.71 |
| 0.16 | 4.67 | 3.23 | 15.10 | 18.33 | 3.71 | 17.30 | 21.01 | 54.16 |
| 0.18 | 5.27 | 3.22 | 16.98 | 20.20 | 3.69 | 19.47 | 23.16 | 77.67 |

**Table SIV Calculated values of traversed fill and flush distances and associated $F_{(L)}$ value for various modulation injection times. 1D flow 0.1 ml/min, 2D flow 20 ml/min, modulation period 7 s.**

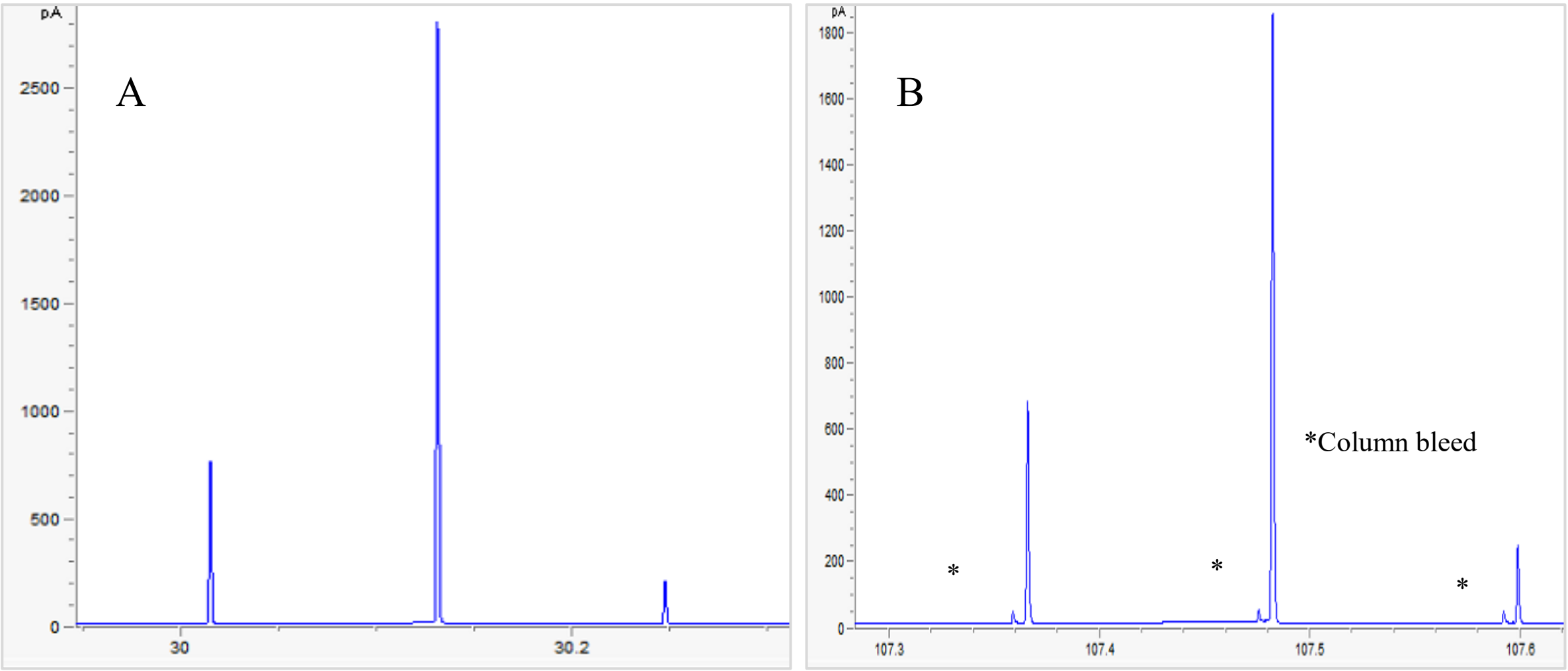

**Figure S5 A) n-C10 modulated peak for injection time 0.12 s; B) n-C28 modulated peak for injection time 0.12 s. 1D flow 0.1 ml/min, 2D flow 20 ml/min, modulation period 7 s.**

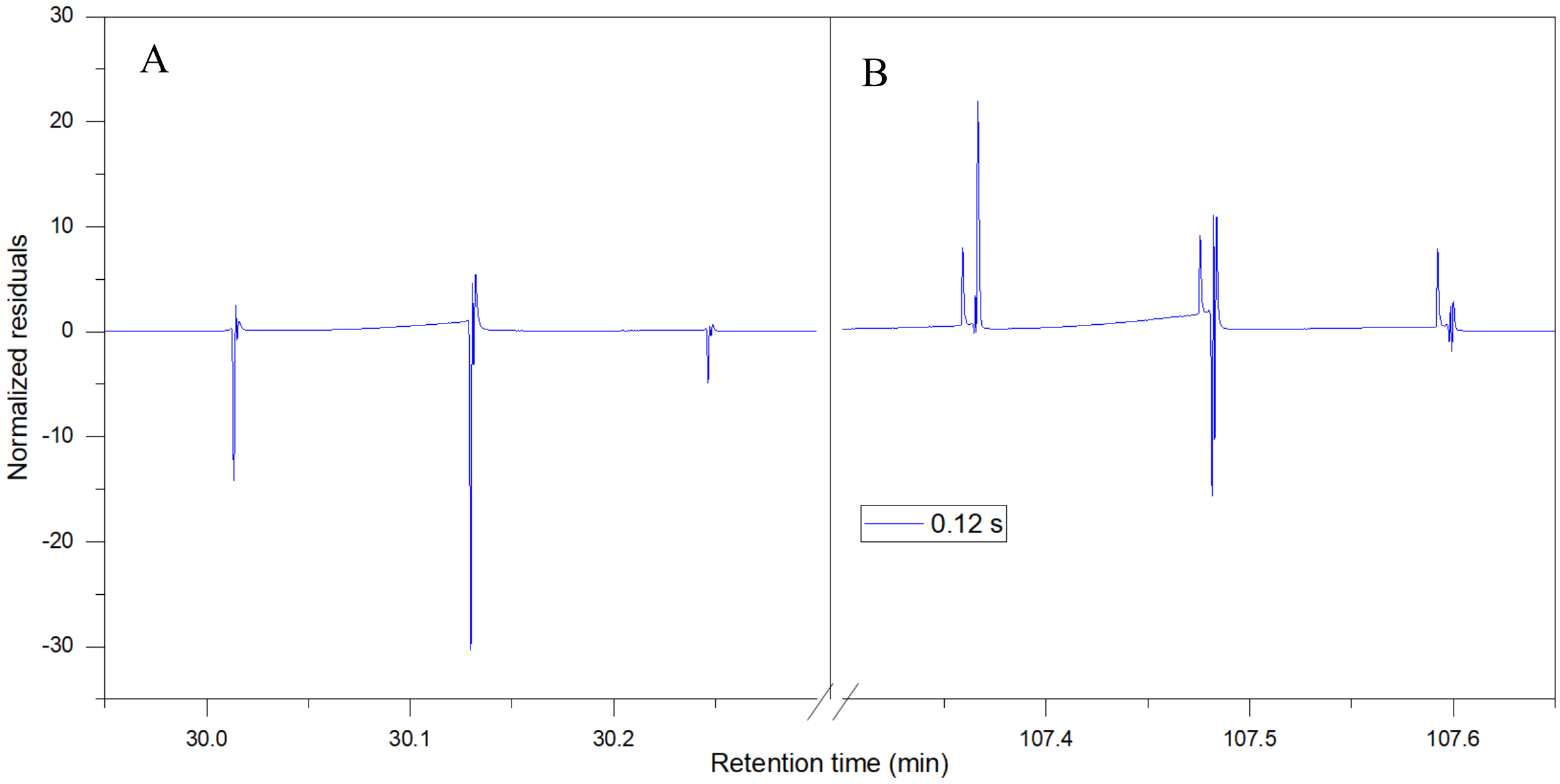

**Figure S6 Normalized residuals for A) n-C10 modulated peak; B) n-C28 modulated peak. 1D flow 0.1 ml/min, 2D flow 20 ml/min, modulation period 7 s, injection time 0.12 s.**

*Set-up **II**: DB-1 (20 m, 0.1 mm ID, 0.4 μm) × BPX-50 (5 m, 0.25 mm ID, 0.25 μm)*
*3.3. Carrier gas: $H_2$*

Second analysis campaign was performed with a DB-1 (20 m×0.1 mm ID×0.4 μm) and BPX-50 (5 m×0.25 mm ID×0.25 μm) columns. This is the set-up we would prefer for flow modulation as longer 2D column decreases the change of the average velocity in the modulation channel from the beginning to the end of the thermally programmed run as previously explained.

As in the case of the previous set-up we can try to determine the best injection time for a run in question. For example, if chosen 1D Flow was 0.1 ml/min, 2D Flow was 8 ml/min and modulation period was 5.5 s, according to the value of $F_{(L)}$, most suitable injection time will be around 0.22-0.25 s, as shown in Table S5.

If we compare chromatograms obtained for injection time 0.25 s and lower injection times, for example 0.18 s, we will observe good modulated peak shape for former case and prominent tailing for the latter indicative of insufficient modulator channel flushing, which will be even more significant for lower injection times (Figure S7). Increase of injection time past the optimum will cause opposite problems in the form of over flushing as previously explained.

| Mod injection time (s) | Flush/Fill | Fill distance (cm) | Flush distance(cm) | Fill +Flush (cm) | Fill distance (cm) | Flush distance(cm) | Fill +Flush (cm) | $F_{(L)}$ |
|---|---|---|---|---|---|---|---|---|
| | | 100°C | | | 300°C | | | |
| 0.14 | 2.09 | 2.98 | 6.22 | 9.20 | 3.52 | 7.35 | 10.88 | 59.00 |
| 0.18 | 2.70 | 2.96 | 7.99 | 10.95 | 3.50 | 9.45 | 12.95 | 35.72 |
| 0.20 | 3.02 | 2.95 | 8.88 | 11.83 | 3.48 | 10.51 | 13.99 | 14.57 |
| 0.22 | 3.33 | 2.94 | 9.77 | 12.71 | 3.47 | 11.56 | 15.03 | 3.47 |
| 0.25 | 3.80 | 2.92 | 11.11 | 14.03 | 3.45 | 13.13 | 16.58 | 3.71 |
| 0.27 | 4.13 | 2.91 | 11.99 | 14.90 | 3.44 | 14.18 | 17.62 | 6.49 |

**Table SV Calculated values of traversed fill and flush distances and associated $F_{(L)}$ value for various modulation injection times. 1D flow 0.1 ml/min, 2D flow 8 ml/min, modulation period 5.5 s.**

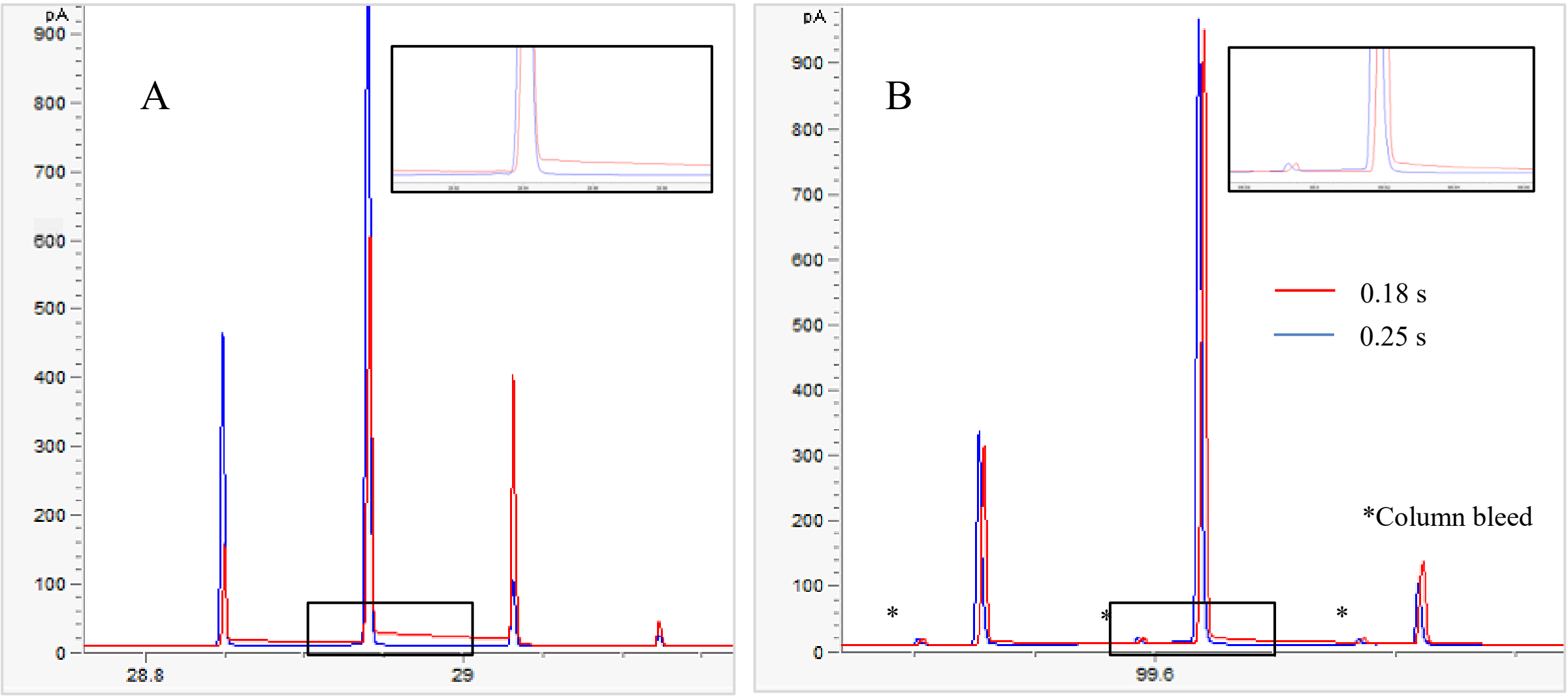

**Figure S7 A) n-C10 modulated peak for injection times 0.18 and 0.25 s; B) n-C28 modulated peak for injection times 0.18 and 0.25 s. 1D flow 0.1 ml/min, 2D flow 8 ml/min, modulation period 5.5 s.**

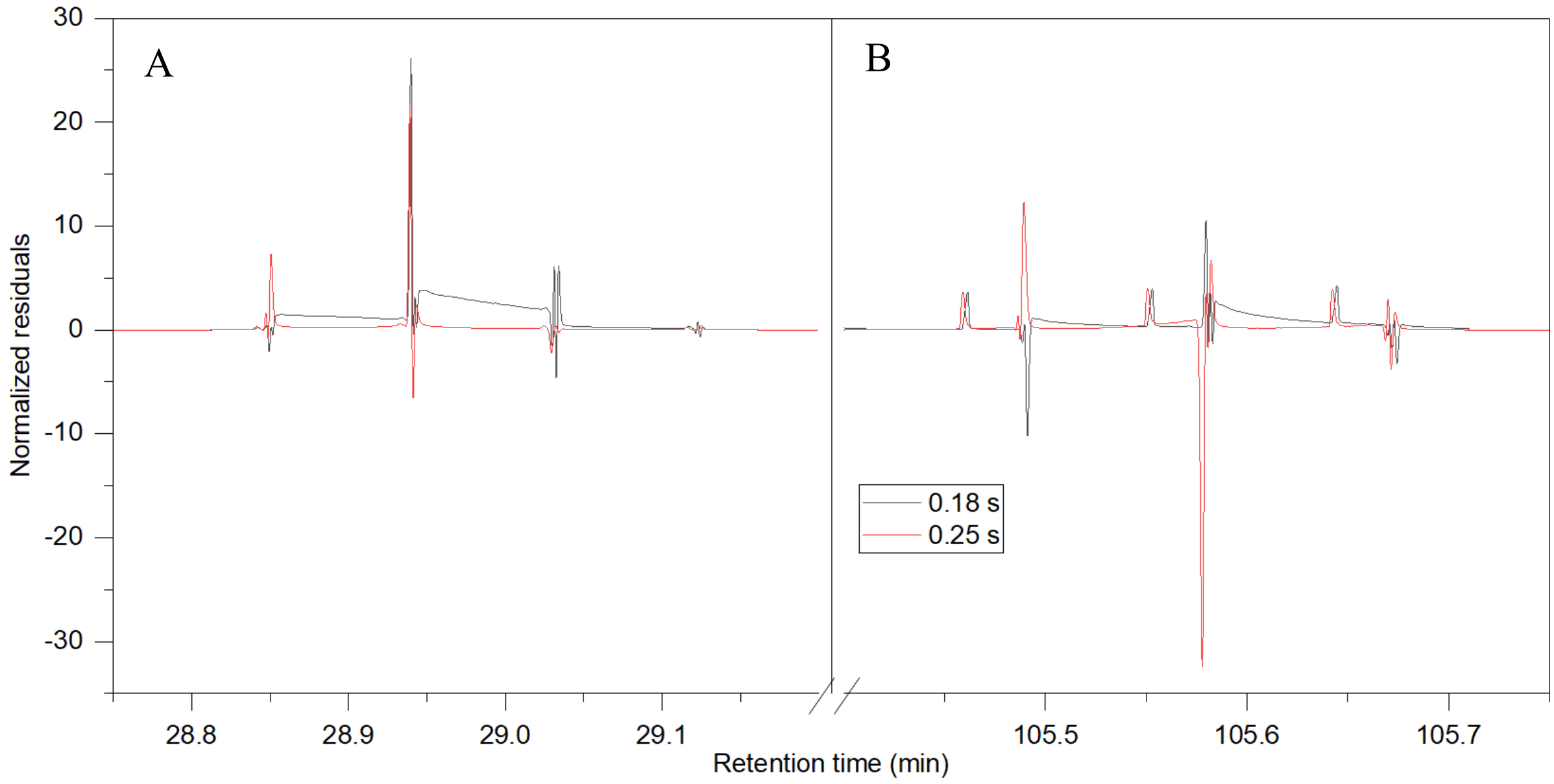

**Figure S8 Normalized residuals for A) n-C10 modulated peak; B) n-C28 modulated peak. 1D flow 0.1 ml/min, 2D flow 8 ml/min, modulation period 5.5 s, injection time 0.18 s and 0.25 s.**

*Set-up **II**: DB-1 (20 m, 0.1 mm ID, 0.4 µm) × BPX-50 (5 m, 0.25 mm ID, 0.25 µm)*
*Carrier gas: He*

In this configuration also carrier gas was changed to He in order to check the prediction performance of the calculations for a different carrier gas.

In the case of 1D Flow 0.15 ml/min, 2D Flow 27 ml/min and modulation period 8 s we have calculated appurtenant fill and flush distances and $F_{(L)}$ values, however this time with He carrier gas. Min F(L) value was obtained for injection time around 0.16-0.18 s (Table S6).

| Mod injection time (s) | Flush/Fill | Fill distance (cm) | Flush distance(cm) | Fill +Flush (cm) | Fill distance (cm) | Flush distance(cm) | Fill +Flush (cm) | $F_{(L)}$ |
|---|---|---|---|---|---|---|---|---|
| | | 100°C | | | 300°C | | | |
| 0.14 | 3.20 | 2.88 | 9.23 | 12.11 | 3.14 | 10.05 | 13.19 | 18.41 |
| 0.16 | 3.67 | 2.87 | 10.54 | 13.41 | 3.13 | 11.49 | 14.62 | 2.58 |
| 0.18 | 4.14 | 2.87 | 11.86 | 14.73 | 3.12 | 12.93 | 16.05 | 1.07 |
| 0.20 | 4.61 | 2.86 | 13.18 | 16.04 | 3.12 | 14.36 | 17.48 | 9.40 |
| 0.22 | 5.08 | 2.85 | 14.50 | 17.35 | 3.11 | 15.80 | 18.91 | 23.33 |
| 0.25 | 5.80 | 2.84 | 16.48 | 19.32 | 3.10 | 17.95 | 21.05 | 59.70 |

**Table SVI Calculated values of traversed fill and flush distances and associated $F_{(L)}$ value for various modulation injection times. 1D flow 0.15 ml/min, 2D flow 27 ml/min, modulation period 8 s.**

Chromatograms obtained for 0.16 and 0.18 s injection times as predicted show very satisfactory peak shapes (Figure S9).

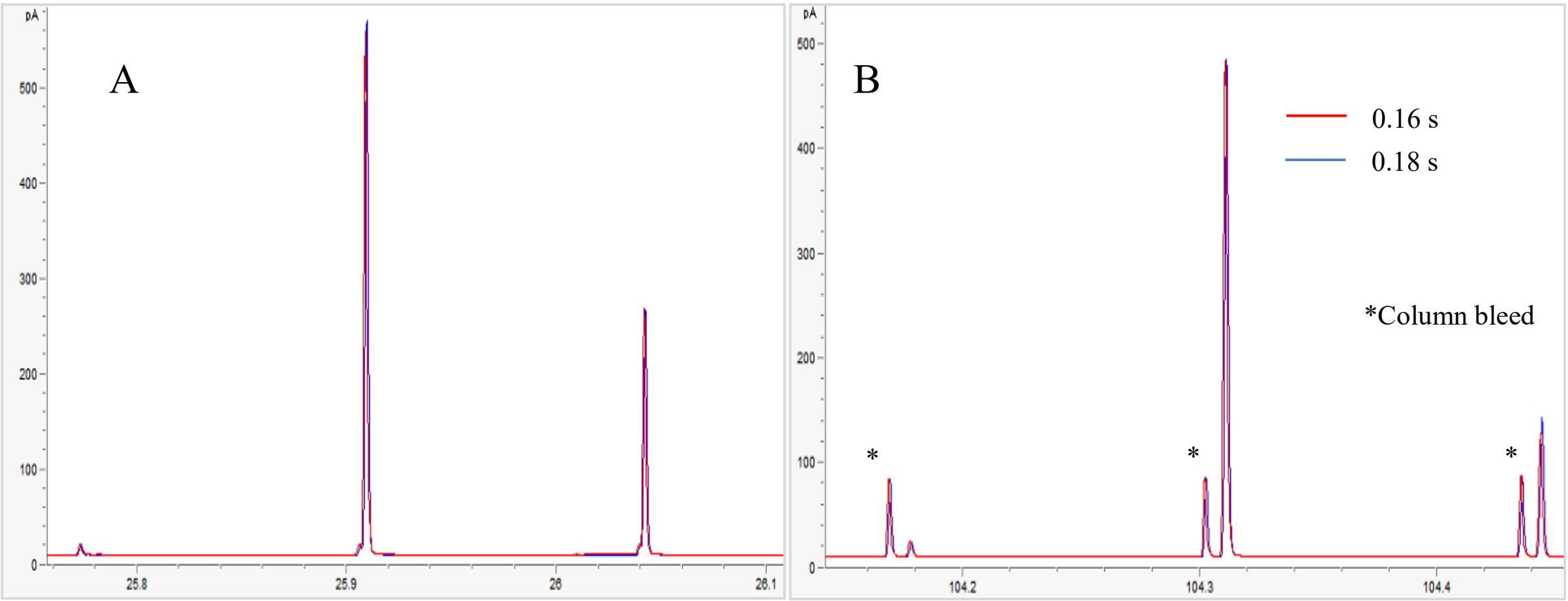

**Figure S9 A) n-C10 modulated peak for injection times 0.16 and 0.18 s; B) n-C28 modulated peak for injection times 0.16 and 0.18 s. 1D flow 0.15 ml/min, 2D flow 27 ml/min, modulation period 8 s.**

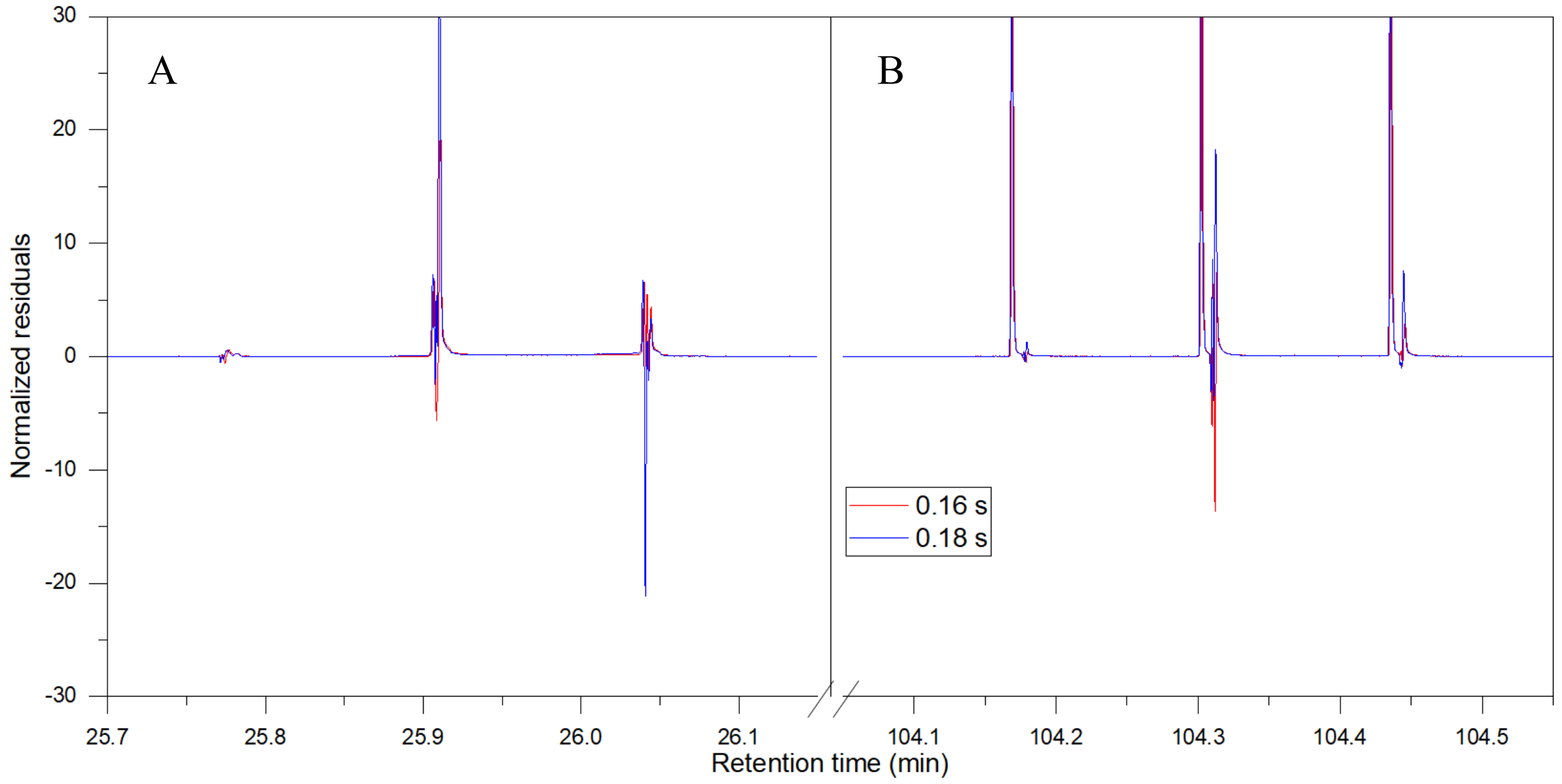

**Figure S10 Normalized residuals for A) n-C10 modulated peak; B) n-C28 modulated peak. 1D flow 0.15 ml/min, 2D flow 27 ml/min, modulation period 8 s, injection time 0.16 s and 0.18 s.**

Also in this configuration with He as carrier gas, in the case of severe under or over-flushing of the modulation channel deteriorating peak shapes will result. For example, in case of 1D Flow 0.15 ml/min, 2D Flow 25 ml/min and modulation period 6 s, 0.3 s injection time will be much too high as shown by our calculations in Table S7.

In line with this, modulated peak shapes both at the beginning and end of chromatogram will be unsatisfactory. In Figure S11 we see severe fronting and double peaks for both n-C10 and n-C28 peak.

| Mod injection time (s) | Flush/Fill | Fill distance (cm) | Flush distance(cm) | Fill +Flush (cm) | Fill distance (cm) | Flush distance(cm) | Fill +Flush (cm) | $F_{(L)}$ |
|---|---|---|---|---|---|---|---|---|
| | | 100°C | | | 300°C | | | |
| 0.3 | 8.76 | 2.16 | 18.97 | 21.31 | 2.36 | 20.70 | 23.06 | 82.72 |

**Table SVII Calculated values of traversed fill and flush distances and associated $F_{(L)}$ value for various modulation injection times. 1D flow 0.15 ml/min, 2D flow 25 ml/min, modulation period 6 s.**

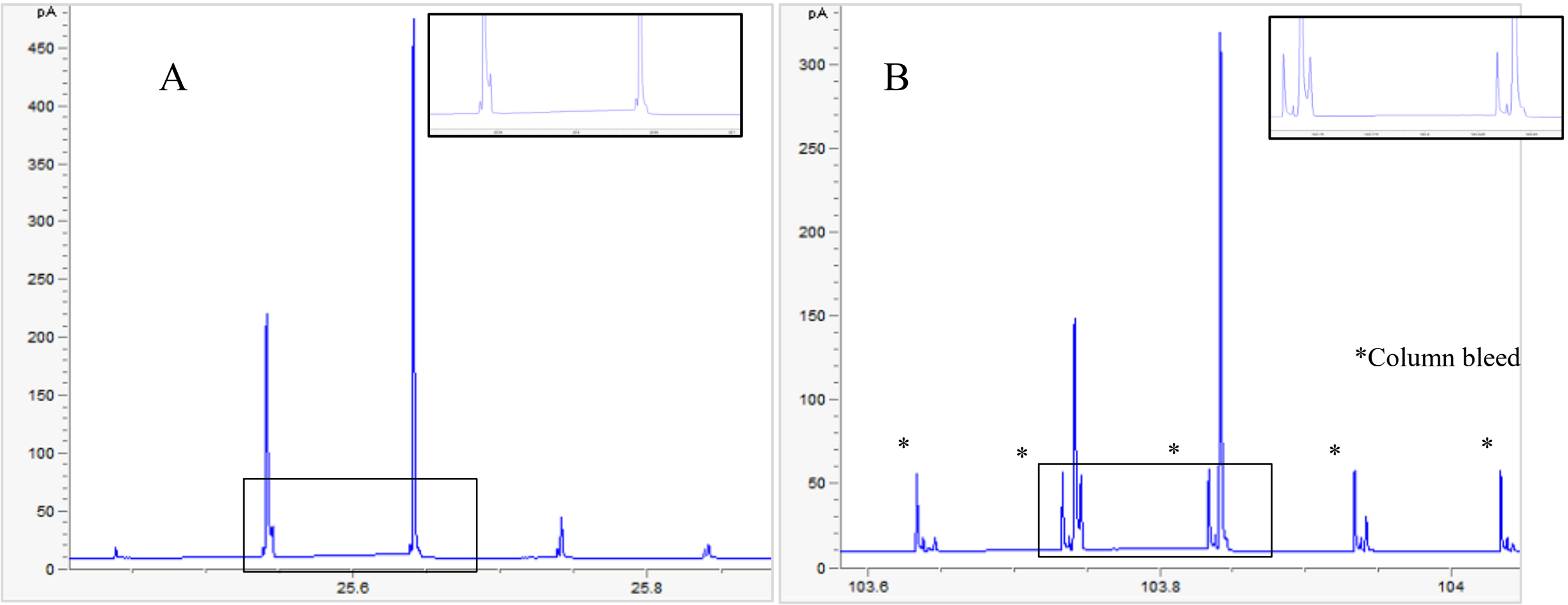

**Figure S11 A) n-C10 modulated peak for injection time 0.30 s; B) n-C28 modulated peak for injection time 0.30 s. 1D flow 0.15 ml/min, 2D flow 25 ml/min, modulation period 6 s.**

*Set-up **III**: Rxi-1ms (30 m, 0.25 mm ID, 0.25 µm) × ZB-35HT (5 m, 0.25 mm ID, 0.18 µm)*
*Carrier gas: $H_2$*

Numerous analysis were also performed with Rxi-1ms (30 m×0.25 mm ID×0.25 µm) and ZB-35HT (5 m×0.25 mm ID×0.18 µm) column. This set up is often employed with forward fill/flush flow modulation, however it can be problematic owing to high 1D flows which are necessarily employed due to higher ID of 1D column, as this will cause flush/fill ratio to be low which might be a compromise efficiency of modulation.

As an example of our prediction for modulation performance, we have taken the analysis in which 1D Flow was 0.3 ml/min, 2D Flow 18 ml/min and modulation period 4 s, in which case we predict the best performance for injection times 0.12-0.14 s (Table S8).

Comparison of chromatograms obtained for 0.10 and 0.12 s injection times indeed demonstrates satisfactory peak shapes for the latter case, while for the former one tailing can be perceived for the modulated peaks both in the beginning and in the end of analysis (Figure S12).

| Mod injection time (s) | Flush/Fill | Fill distance (cm) | Flush distance(cm) | Fill +Flush (cm) | Fill distance (cm) | Flush distance(cm) | Fill +Flush (cm) | $F_{(L)}$ |
|---|---|---|---|---|---|---|---|---|
| | | 100°C | | | 300°C | | | |
| 0.10 | 1.54 | 4.85 | 7.45 | 12.30 | 5.46 | 8.39 | 13.85 | 14.76 |
| 0.12 | 1.85 | 4.82 | 8.94 | 13.76 | 5.43 | 10.07 | 15.50 | 3.06 |
| 0.14 | 2.17 | 4.79 | 10.43 | 15.22 | 5.40 | 11.75 | 17.15 | 4.16 |
| 0.16 | 2.50 | 4.77 | 11.92 | 16.69 | 5.37 | 13.43 | 18.80 | 20.77 |

**Table SVIII Calculated values of traversed fill and flush distances and associated $F_{(L)}$ value for various modulation injection times. 1D flow 0.3 ml/min, 2D flow 18 ml/min, modulation period 4 s.**

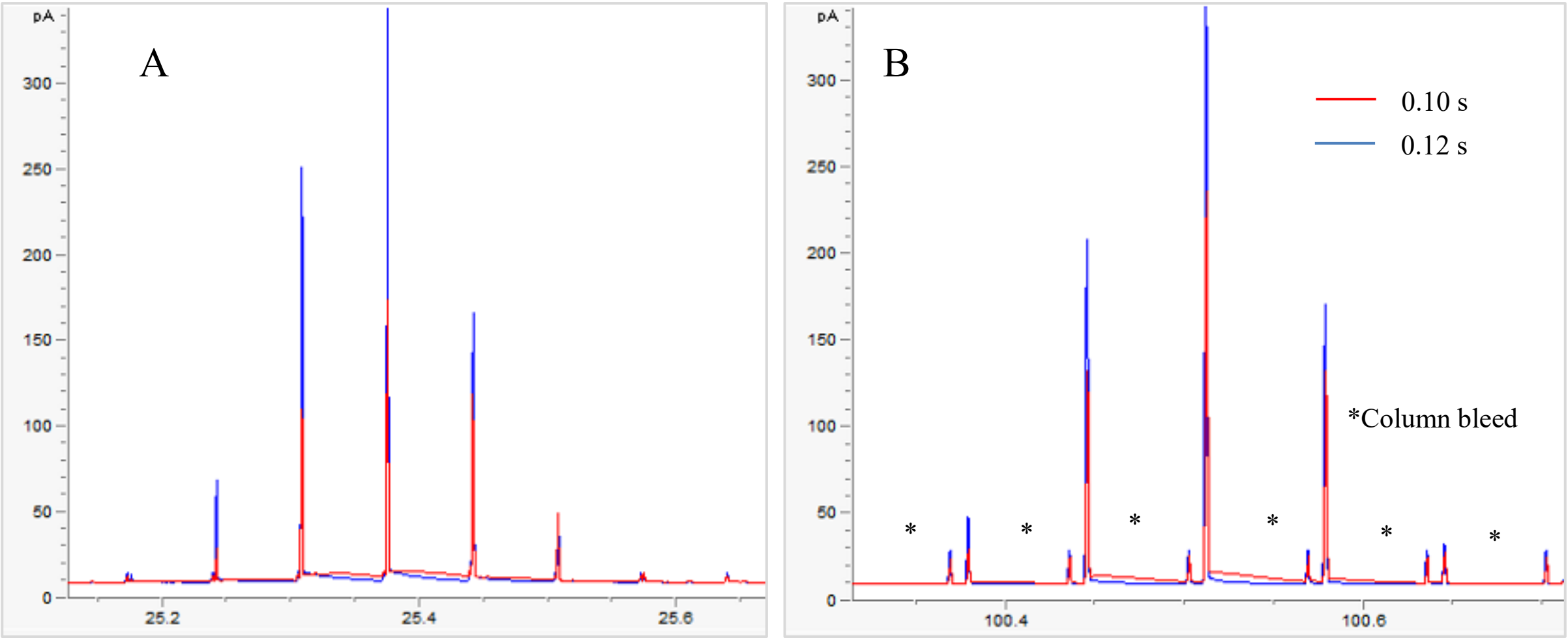

**Figure S12 A) n-C10 modulated peak for injection times 0.10 and 0.12 s; B) n-C28 modulated peak for injection times 0.10 and 0.12 s. 1D flow 0.3 ml/min, 2D flow 18 ml/min, modulation period 4 s.**

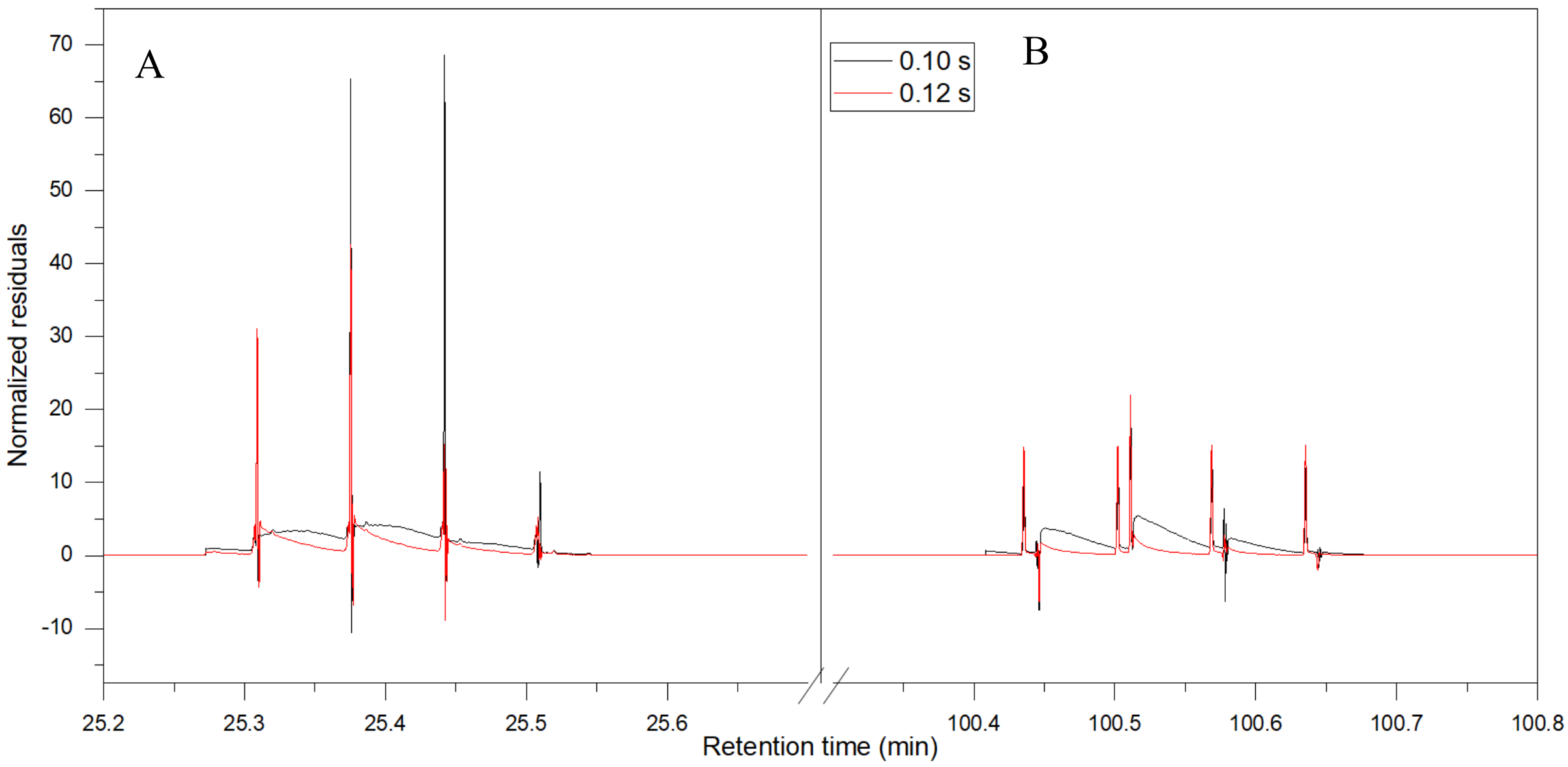

**Figure S13 Normalized residuals for A) n-C10 modulated peak; B) n-C28 modulated peak. 1D flow 0.3 ml/min, 2D flow 18 ml/min, modulation period 4 s, injection time 0.10 s and 0.12 s.**

As previously mentioned it is important to note that changing of oven temperature ramp can also affect the modulation of peaks. Since in GC×GC separations oven ramps are rather low we can assume that temperature is almost the same for the modulation over 1D peak. To verify this, we have compared modulated peak shape in the same run in which we have only changed oven ramp.

We have employed 1D Flow 0.3 ml/min, 2D Flow 27 ml/min, modulation period 6 s and injection time of 0.1 s which we have calculated to be optimal for this analysis (Table S9).

We have acquired chromatogram at oven ramps: 1.5, 2.5 and 3°C/min and compared modulated peaks over chromatograms (Figure S14). We have not observed any significant difference in the shapes of the peaks, which testifies that our prediction can be safely employed for usual GC×GC oven ramps.

| Mod injection time (s) | Flush/Fill | Fill distance (cm) | Flush distance(cm) | Fill +Flush (cm) | Fill distance (cm) | Flush distance(cm) | Fill +Flush (cm) | $F_{(L)}$ |
|---|---|---|---|---|---|---|---|---|
| | | 100°C | | | 300°C | | | |
| 0.10 | 1.52 | 6.19 | 9.44 | 15.63 | 6.87 | 10.47 | 17.33 | 6.3 |

**Table SIX Calculated values of traversed fill and flush distances and associated $F_{(L)}$ value for 1D flow 0.3 ml/min, 2D flow 27 ml/min, modulation period 6 s and injection time 0.1 s.**

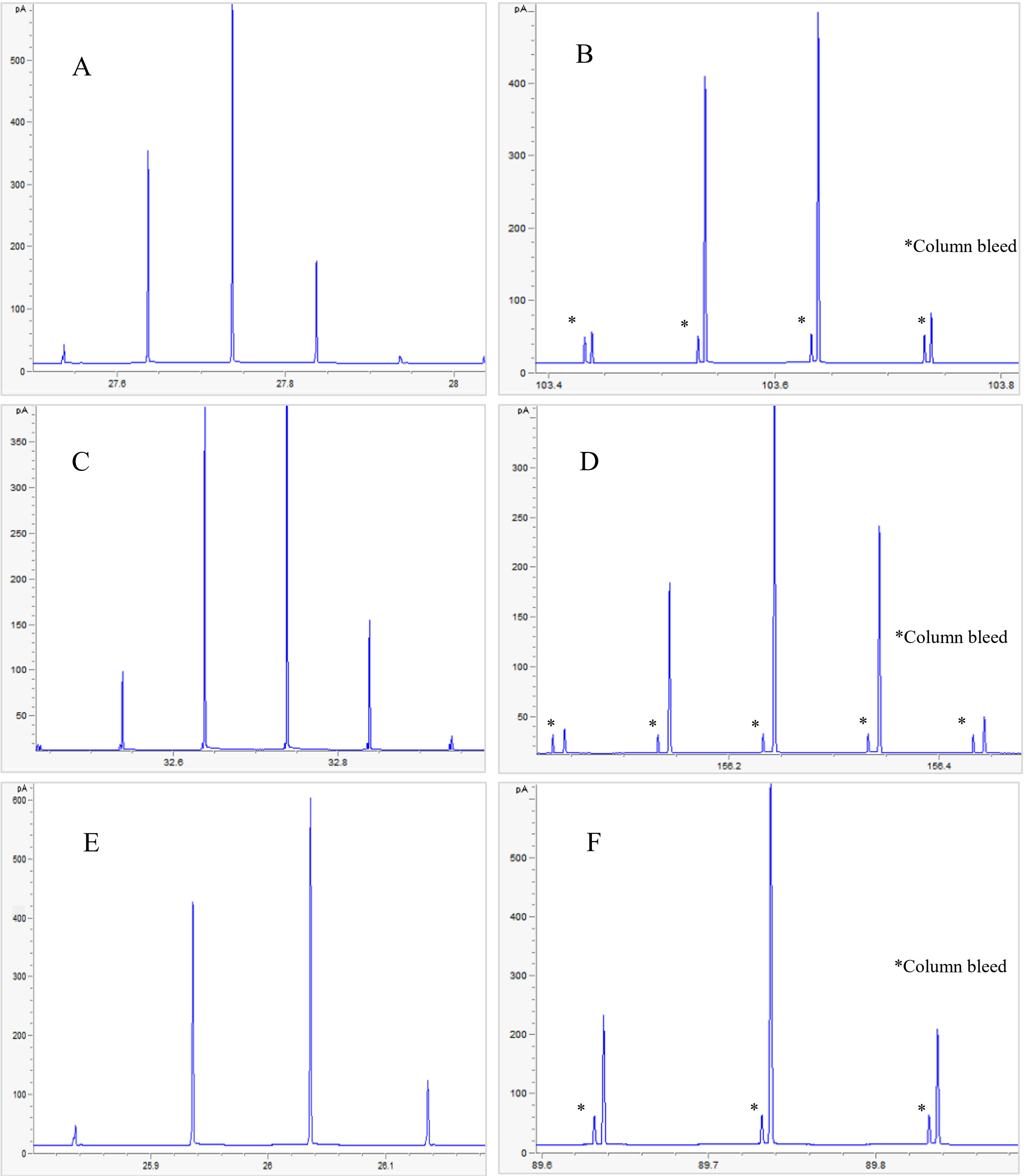

**Figure S14 A) n-C10 modulated peak for oven ramp 2.5 °C/min; B) n-C28 modulated peak for oven ramp 2.5 °C/min; C) n-C10 modulated peak for oven ramp 1.5 °C/min; D) n-C28 modulated peak for oven ramp 1.5 °C/min.; E) n-C10 modulated peak for oven ramp 3 °C/min; F) n-C28 modulated peak for oven ramp 3 °C/min. 1D flow 0.3 ml/min, 2D flow 27 ml/min, modulation period 6 s and injection time 0.1 s.**

*Set-up IV: ZB-5HT (15 m, 0.1 mm ID, 0.1 µm) × ZB-35HT (5 m, 0.25 mm ID, 0.18 µm)*
*Carrier gas: $H_2$*

Finally, analysis were also performed with ZB-5HT (15 m×0.1 mm ID×0.1 µm) and ZB-35HT (5 m×0.25 mm ID×0.25 µm) columns, which is a set-up adapted for the analysis of samples containing very high boiling point analytes. Hence, we have employed Test Mix with wider range of carbon number, from n-C14 to n-C44. Since 1D column in this case has a thin stationary phase, it is very easily overloaded with sample and 1D peaks can easily demonstrate significant fronting.

For demonstration, we used 1D Flow 0.12 ml/min, 2D Flow 12 ml/min and modulation period 7 s. The best injection time according to our calculations would be around 0.16-0.18 s (Table S10).
Chromatograms for the best predicted injection times are shown in Figure S15. Peak shapes are good. As in all previous cases lower injection times will result in tailing, which can be seen on the chromatogram acquired for 0.14 s for n-C14 peak (Figure S16A). Higher injection times will result in fronting, which is shown for 0.24 s injection time for n-C44 peak (Figure S16B).

| Mod injection time (s) | Flush/Fill | Fill distance (cm) | Flush distance(cm) | Fill +Flush (cm) | Fill distance (cm) | Flush distance(cm) | Fill +Flush (cm) | $F_{(L)}$ |
|---|---|---|---|---|---|---|---|---|
| | | 150 °C | | | 350 °C | | | |
| 0.14 | 2.04 | 4.17 | 8.49 | 12.66 | 4.69 | 9.56 | 14.25 | 6.40 |
| 0.16 | 2.34 | 4.16 | 9.71 | 13.87 | 4.68 | 10.93 | 15.60 | 2.79 |
| 0.18 | 2.64 | 4.14 | 10.93 | 15.07 | 4.66 | 12.29 | 16.95 | 3.23 |
| 0.20 | 2.94 | 4.13 | 12.14 | 16.27 | 4.65 | 13.66 | 18.31 | 16.59 |
| 0.22 | 3.24 | 4.12 | 13.36 | 17.48 | 4.63 | 15.02 | 19.66 | 27.71 |
| 0.24 | 3.55 | 4.11 | 14.57 | 18.68 | 4.62 | 16.39 | 21.69 | 56.85 |
| 0.26 | 3.85 | 4.10 | 15.78 | 19.88 | 4.62 | 16.39 | 21.01 | 72.02 |

**Table SX Calculated values of traversed fill and flush distances and associated $F_{(L)}$ value for various modulation injection times. 1D flow 0.12 ml/min, 2D flow 12 ml/min, modulation period 7 s.**

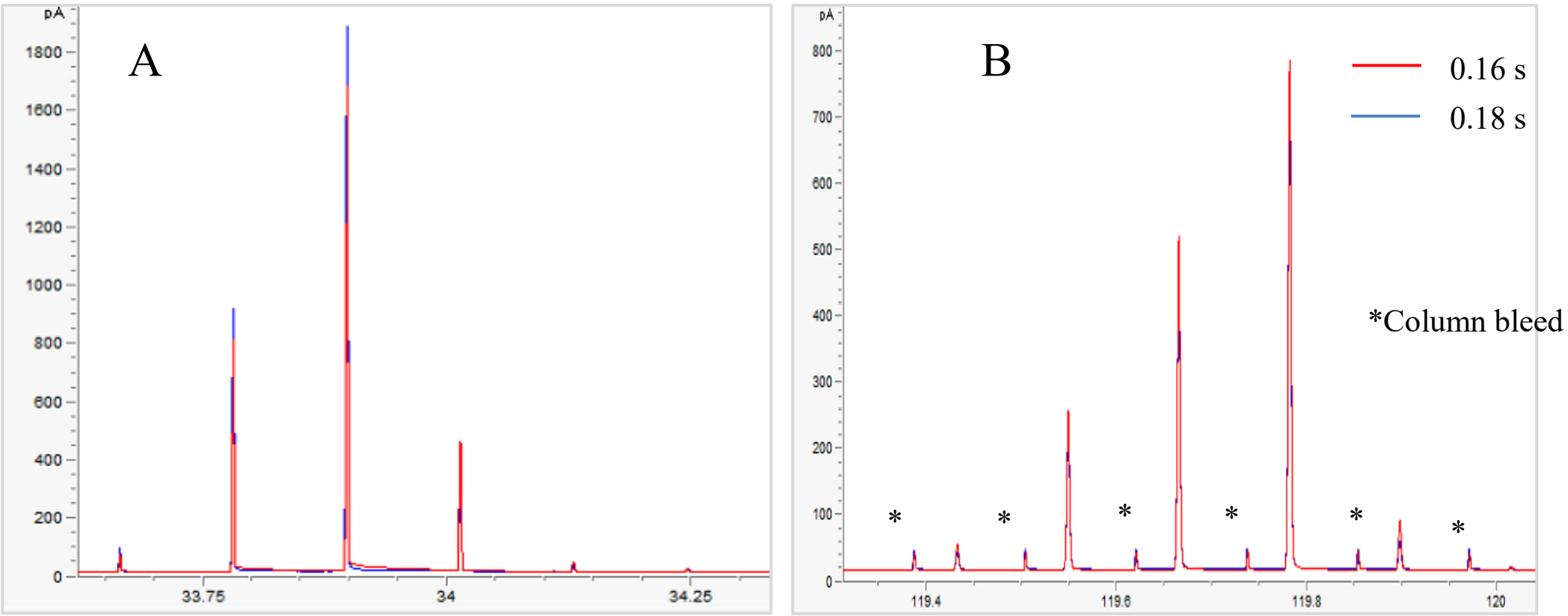

**Figure S15 A) n-C14 modulated peak for injection times 0.16 and 0.18 s; B) n-C44 modulated peak for injection times 0.16 and 0.18 s. 1D flow 0.12 ml/min, 2D flow 12 ml/min, modulation period 7 s.**

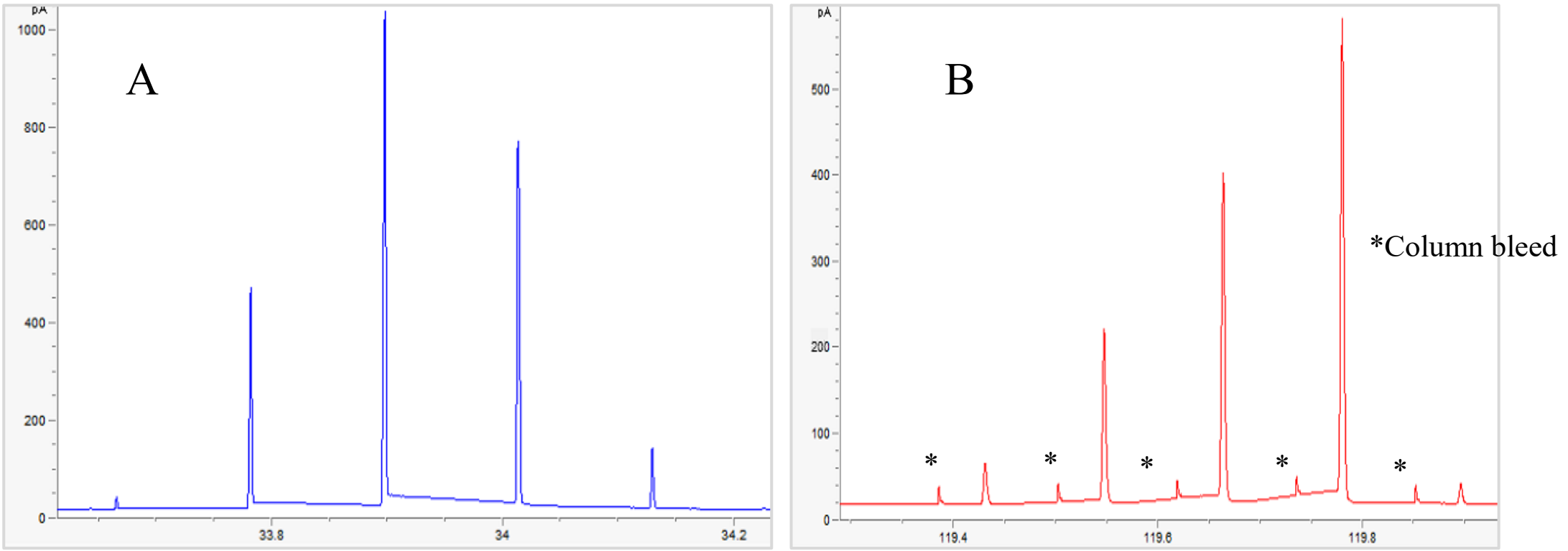

**Figure S16 A) n-C14 modulated peak for injection time 0.14 s; B) n-C44 modulated peak for injection time 0.24 s. 1D flow 0.12 ml/min, 2D flow 12 ml/min, modulation period 7 s.**

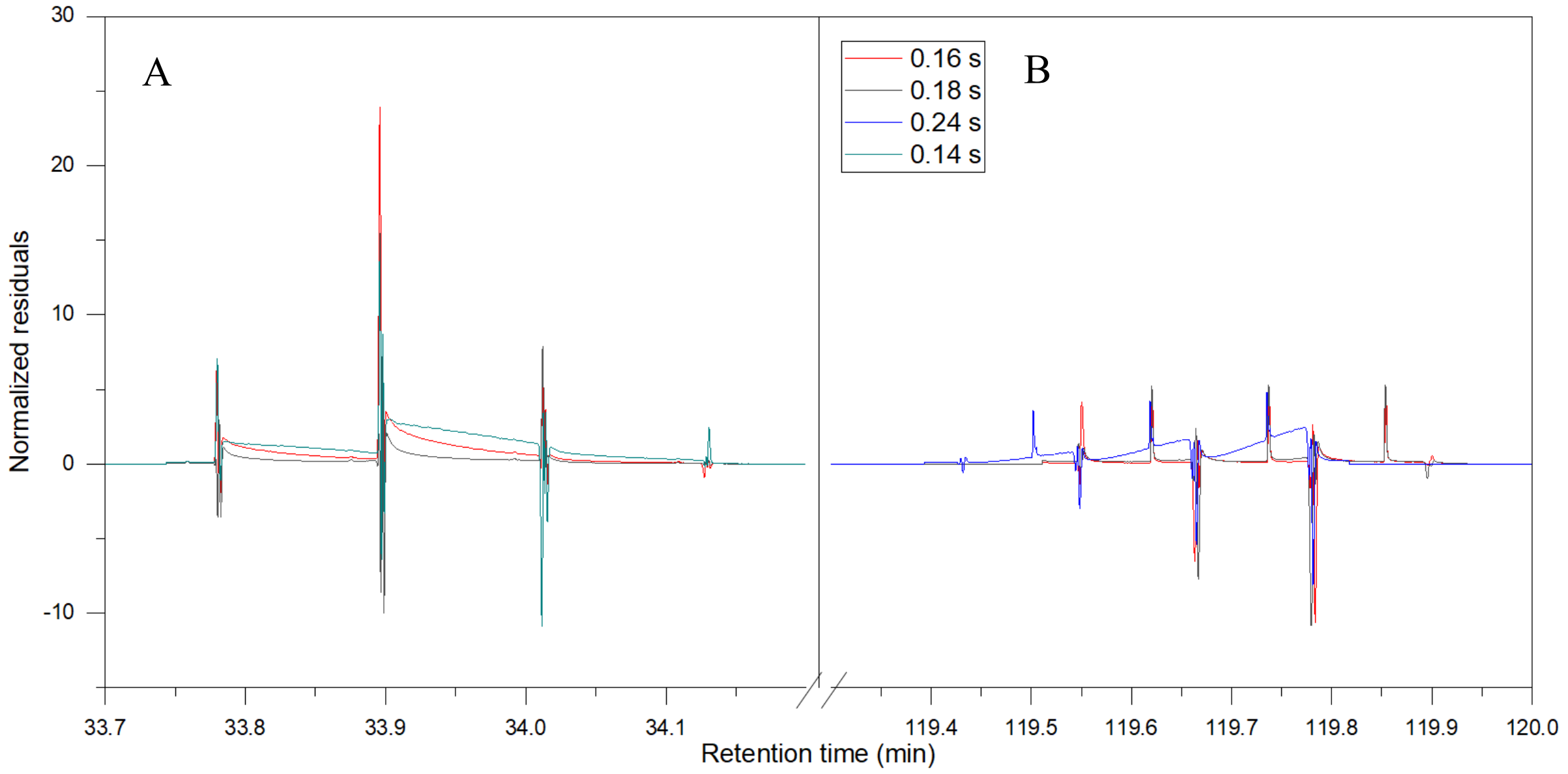

**Figure S17 Normalized residuals for A) n-C10 modulated peak; B) n-C28 modulated peak. 1D flow 0.12 ml/min, 2D flow 12 ml/min, modulation period 7 s, injection time 0.14 s to 0.24 s.**

*Modulation tests with a more complex test mixture.*

00.02.718 PNA in Diesel - Gravimetric blend from AC Analytical Controls® (PAC) was analysed with following GC×GC analysis conditions: DB-1 (20 m, 0.1 mm ID, 0.4 µm) and ZB-50 (10 m, 0.25 mm ID, 0.1 µm) column set, $^1$D flow 0.25 ml/min and $^2$D flow 15 ml/min, modulation period 5 s. Oven programming was 80 °C (1min) to 325 °C at 2 °C/min. Carrier gas was hydrogen.

Best injection time calculated according to the Flow modulation calculator was 0.19 s with F(L)=3.7 and Flush/Fill=2.37. Fill +Flush (cm) at 80 °C was 13.35 cm and Fill +Flush (cm) at 325 °C was 15.10 cm. Obtained 2D chromatogram for 0.19 s injection time is provided in Figure S18. Good peak shape was obtained for all compounds (n-paraffins, naphthenes, FAMEs, mono and diaromatics). 1D chromatogram preview is provided in Figure S19. It can be seen that very good beak shape with minimal tailing or fronting was generated for n-C30 (Fill +Flush ~ 15.10 cm). For light analytes also very good peak shape is obtained with minimal tailing which is expected as Fill +Flush ~ 13.35 cm.

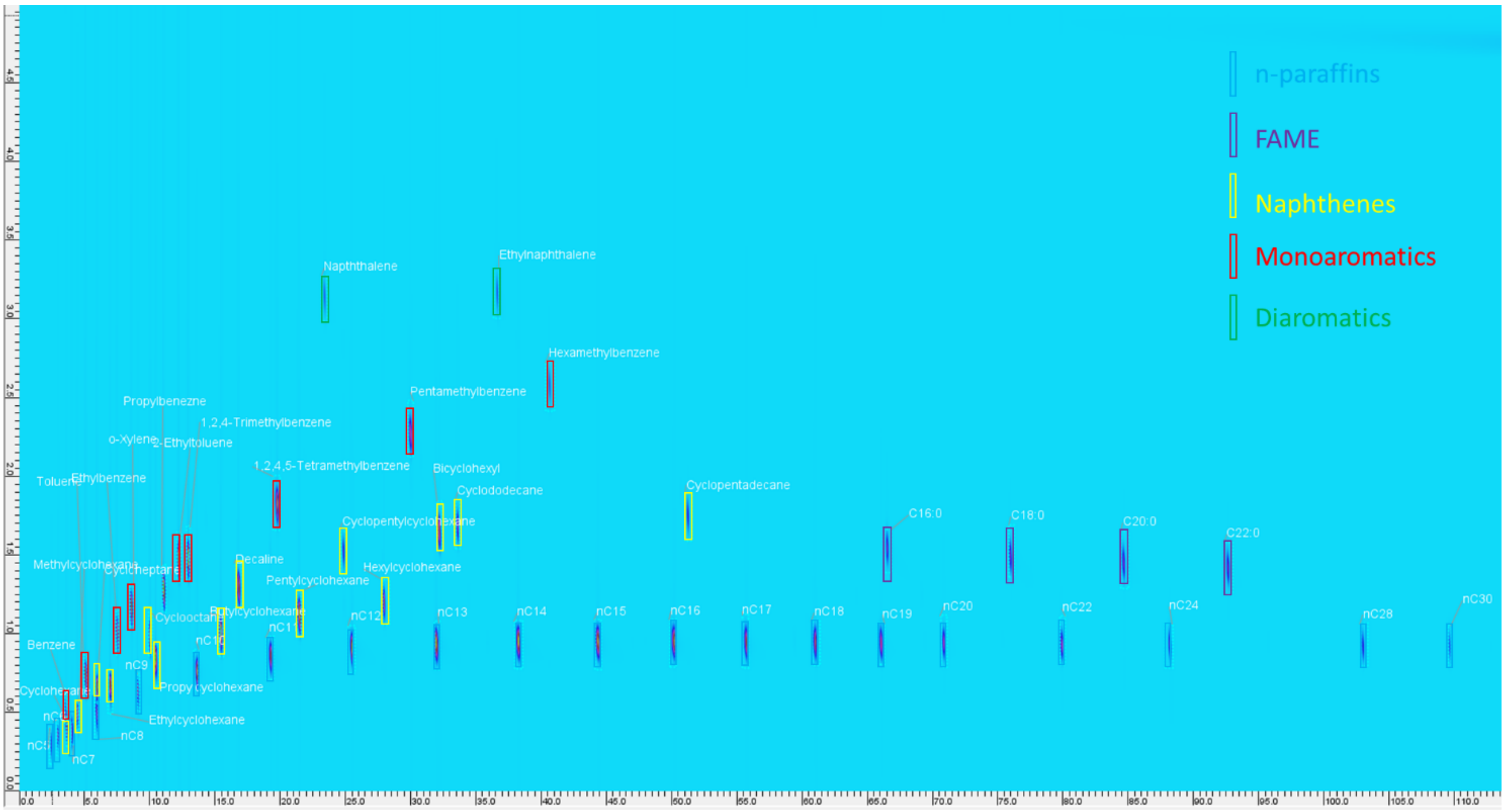

**Figure S18 2D chromatogram of a commercial gravimetric test mix obtained for modulation injection time of 0.19 s.**

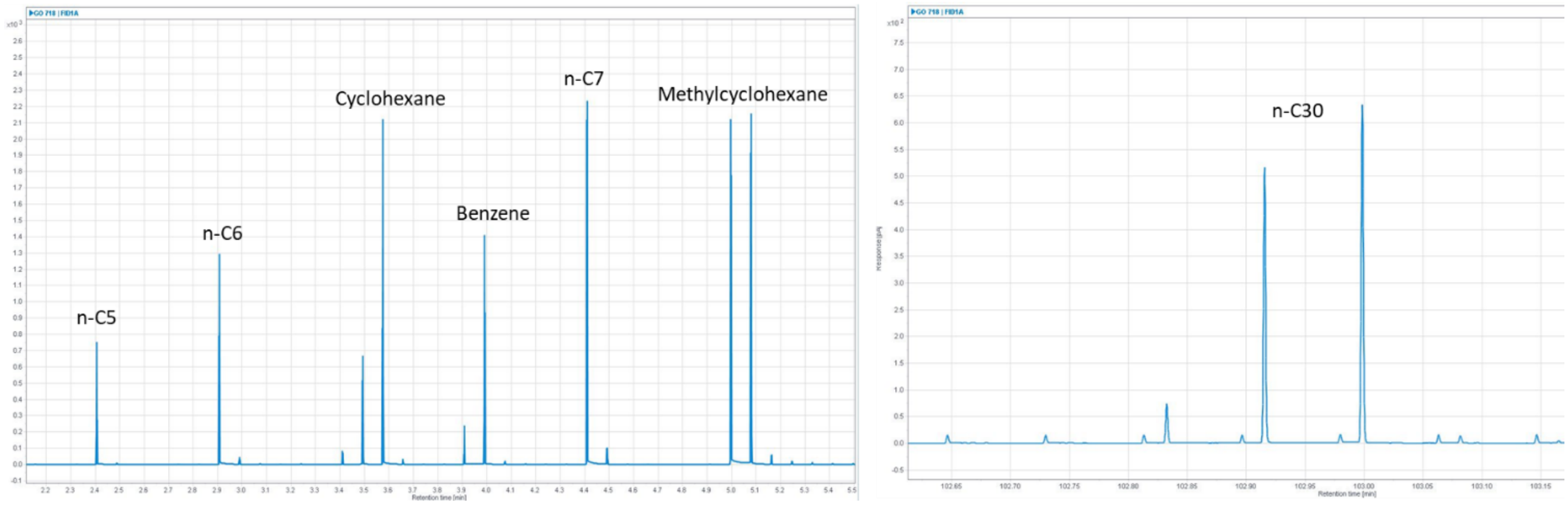

**Figure S19 1D chromatogram of a commercial gravimetric test mix obtained for modulation injection time of 0.19 s (zoom on the zones of light and heavy analytes).**

If we try to increase the modulation time to test the performance of Flow modulation calculator, for 0.3 s injection time we will obtain F(L) = 57.63 and for 0.4 s F(L) = 118.67.

We can see in the 2D chromatograms (acquired for injection times 0.3 s and 0.4 s) in Figure S20 that peak shape is seriously degraded compared to injection of 0.19 s, which is predicted by the F(L) values. Such injection times cause over-flushing of the modulation channel. It can be seen also that peak shape does not depend on the type of compound and that all analytes' peak shapes degrade in exactly the same way. This shows that behaviour of peak shapes for all analysis can be approximated by analysing a simple test mixture of n-paraffins as preformed in this work.

Zoom on the 1D chromatogram regions of light and heavy analytes for injection times 0.19 s, 0.3 s and 0.4 s is provided in Figure S21. With the increase of the injection time, peak shapes degrade for all analytes, resulting even in double peaks for 0.4 s injection (see n-C30 modulated peak).

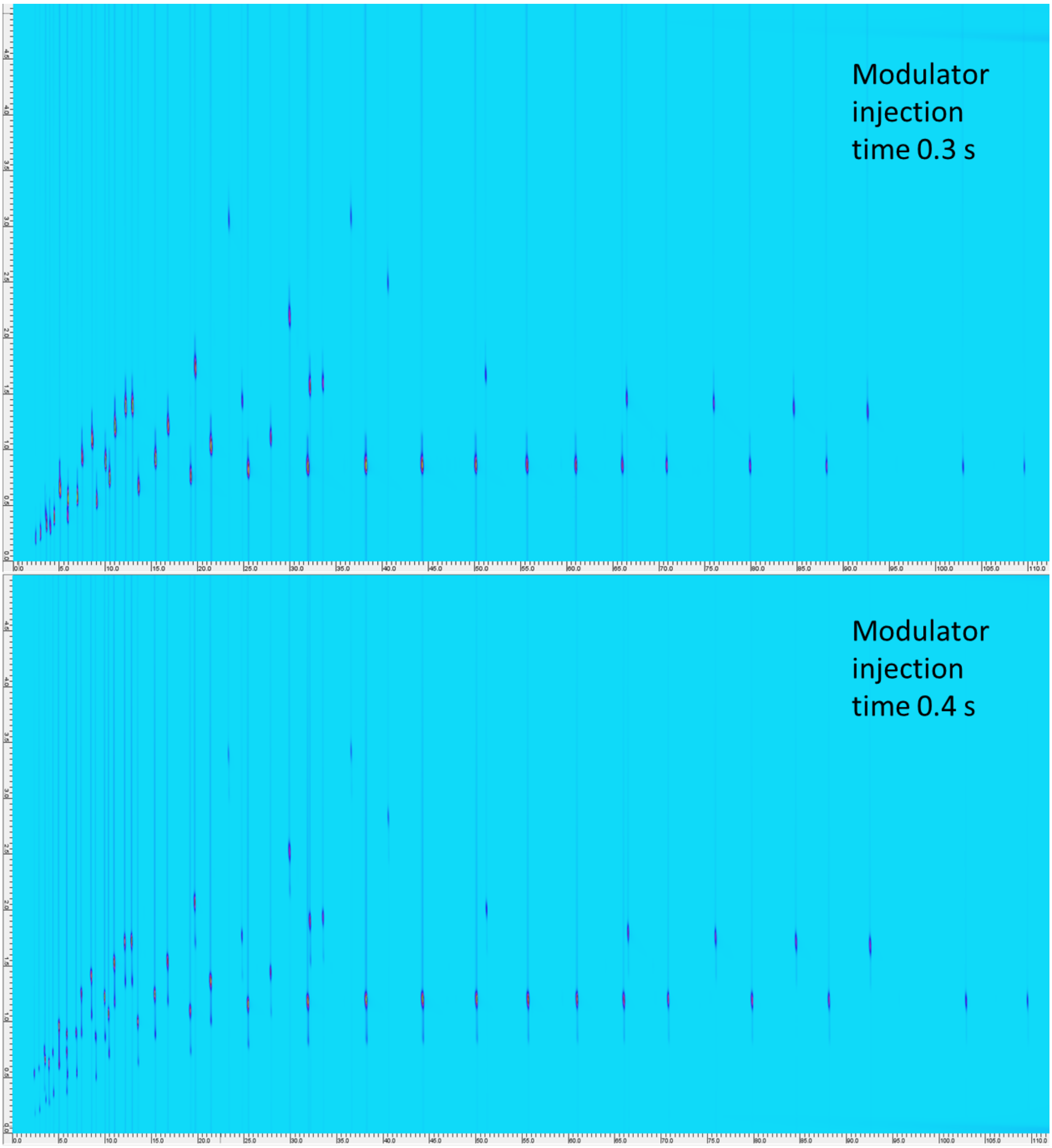

**Figure S20 2D chromatogram of a commercial gravimetric test mix obtained for modulation injection time of 0.19 s and 0.3 s.**

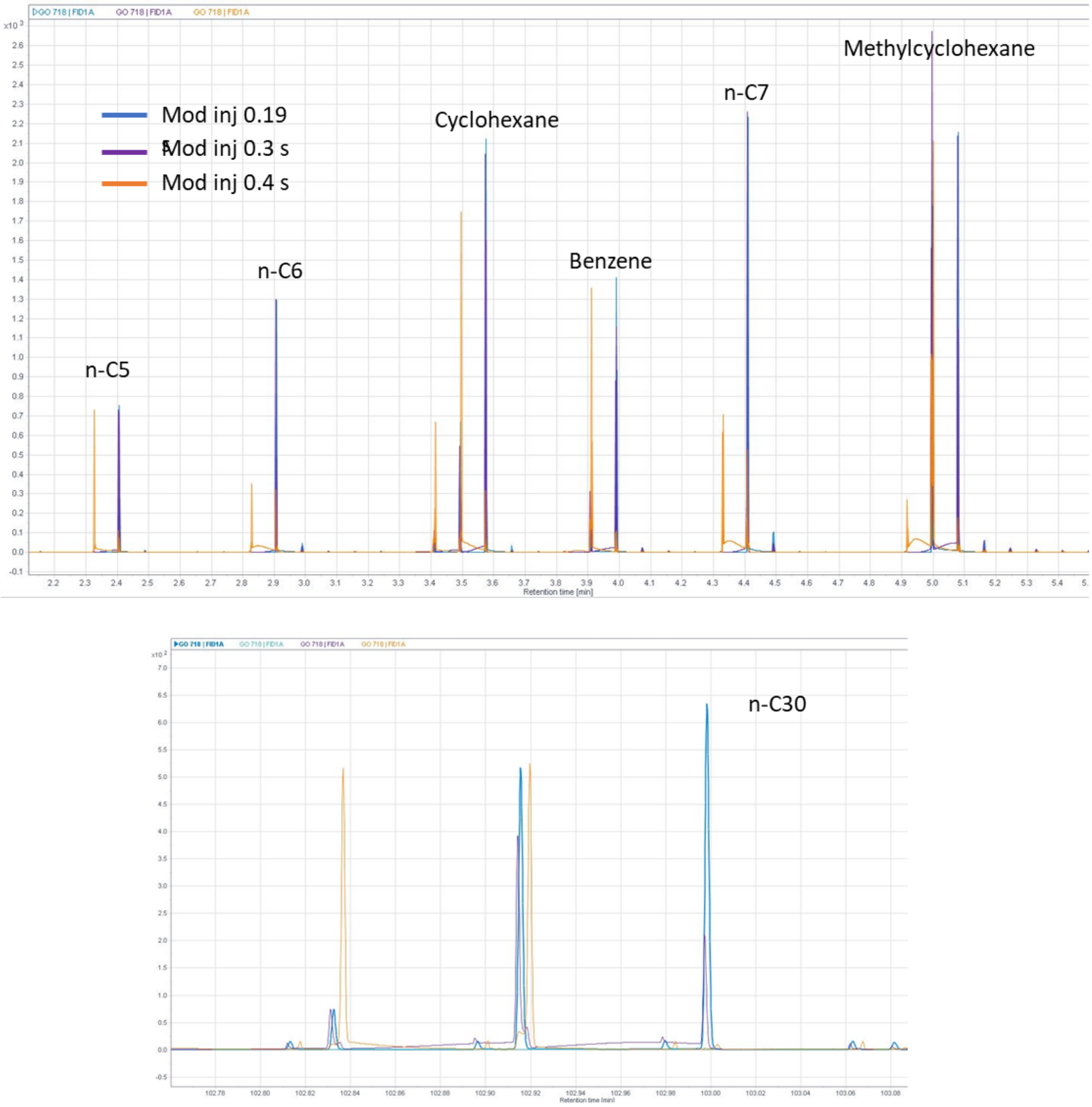

**Figure S21 1D chromatogram of a commercial gravimetric test mix obtained for modulation injection time of 0.19, 0.3 s and 0.4 s (zoom on the zones of light and heavy analytes).**

Flow-modulation-calculator-Code-new.py

```python
import sys
import numpy as np
import tkinter as tk
from tkinter import *
import math

fields = ('1D Length* (m)', '1D Diameter* (mm)', '2D Length* (m)', '2D Diameter* (mm)','Initial temperature* (°C)','Final temperature* (°C)','2D outlet pressure* (psi)','1D Flow* (ml/min)','2D Flow (ml/min)','Modulation period (s)','Injection time (s)','Function','Flush/Fill')


def calculate(entries):

    L1= float(entries['1D Length* (m)'].get())
    L2 = float(entries['2D Length* (m)'].get())
    d1 =  float(entries['1D Diameter* (mm)'].get())
    d2 =  float(entries['2D Diameter* (mm)'].get())
    F1= float(entries['1D Flow* (ml/min)'].get())
    F2 = float(entries['2D Flow (ml/min)'].get())
    if F2==0:
        F2 = [6,7,8,9,10,11,12,13,14,15,16,17,18,19,20,21,22,23,24,25,26,27]
    P = float(entries['Modulation period (s)'].get())
    if P==0:
        P = [1,2,3,4,5,6,7,8,9]
    t = float(entries['Injection time (s)'].get())
    if t==0:
        t = [0.10,0.11,0.12,0.13,0.14,0.15,0.16,0.17,0.18,0.19,0.20,0.21,0.22,0.23,0.24,0.25,0.26,0.27,0.28,0.30,0.31,0.32,0.33,0.34,0.35,0.36,0.37,0.38,0.39,0.4,0.41,0.42,0.43,0.44,0.45,0.46,0.47,0.48,0.49,0.50]
    T0 = float(entries['Initial temperature* (°C)'].get())
    Tf = float(entries['Final temperature* (°C)'].get())
    T=[T0,Tf]
    Pout = float(entries['2D outlet pressure* (psi)'].get())
    gas=str(carrierGas.get())
    fullresult=[]
    FunctList=[]
    Elements=[]


    #constants definition:
    dmod=0.535
    Lmod=0.15
    Tref=25
    Pref=14.696
    Pconv=6894.757
    Vconv=1.66666666666667*10**(-8)
    pi=math.pi
    #viscosity at standard temperature:
    VHe=18.69*10**(-6)
    VH2 = 8.362 * 10 ** (-6)
    VN2 = 16.24 * 10 ** (-6)
    VAr = 21.35 * 10 ** (-6)
    #gas dependent exponent:
    eHe=0.685
    eH2 = 0.698
    eN2 = 0.710
    eAr = 0.750

    #estimate Combinations:
    np.set_printoptions(threshold=sys.maxsize)
    new_array = np.array(
        np.meshgrid(L1, d1, L2, d2, F1, F2, P, t)).T.reshape(-1, 8)
    #print(new_array)
    for i in new_array:
        L1=i[0]
        d1= i[1]
        L2 = i[2]
        d2 = i[3]
        F1 = i[4]
        F2 = i[5]
        P = i[6]
        t=i[7]
        result = []
        elem=[]
        for j in range(len(T)):
            Tj=T[j]
            #Calculation of viscosities:
            if gas=='Hydrogen':
                Visc=(((Tj+273.15)/273.15)**eH2)*VH2
            elif gas=='Nitrogen':
                Visc = (((Tj + 273.15) / 273.15) ** eN2) * VN2
            elif gas == 'Helium':
                Visc = (((Tj + 273.15) / 273.15) ** eHe) * VHe
            elif gas == 'Argon':
                Visc = (((Tj + 273.15) / 273.15) ** eAr) * VAr
            # Calculation of column constants:
            K1D=((pi*(d1/2)**4*10**(-12))/(16*Visc*L1))*(1/(Pref*Pconv))*((Tref+273.15))/(Tj+273.15)
            K2D=((pi*(d2/2)**4*10**(-12))/(16*Visc*L2))*(1/(Pref*Pconv))*((Tref+273.15))/(Tj+273.15)
            Kmod=((pi*(dmod/2)**4*10**(-12))/(16*Visc*Lmod))*(1/(Pref*Pconv))*((Tref+273.15))/(Tj+273.15)
            #Calculation of pressures:
            P2Din = math.sqrt((F2 * Vconv / K2D) + (Pout * Pconv) ** 2) / Pconv
            P1Din=math.sqrt((F1*Vconv/K1D)+(P2Din*Pconv)**2)/Pconv
            #Fill mode pressure and flow for the modulator:
            Pmodout1=P2Din
            Pmodin1=math.sqrt(((d1**4*Lmod*P1Din**2)+(dmod**4*L1*P2Din**2))/((dmod**4*L1)+(d1**4*Lmod)))
            Fmod1=Kmod*((Pmodin1*Pconv)**2-(Pmodout1*Pconv)**2)/Vconv
            #Fill mode average velocity calculation:
            Pratio1=Pmodin1/Pmodout1
            C1=3/2*((Pratio1**2-1)/(Pratio1**3-1))
            Vmod1=pi*(dmod/2000)**2*Lmod
            Veloc1=Pref*Pconv*(Tj+273.15)*Lmod*Fmod1*Vconv*C1/(Pmodout1*Pconv*(Tref+273.15)*Vmod1)*100
            Fill=Veloc1*(P-t)
            # Flush mode pressure and flow for the modulator:
            Pmodin2 = P2Din
            Pmodout2 = math.sqrt(((dmod**4*L2*P2Din**2) + (d2**4*Lmod*Pout**2))/((d2**4*Lmod)+(dmod**4*L2)))
            Fmod2 = Kmod * ((Pmodin2 * Pconv) ** 2 - (Pmodout2 * Pconv) ** 2) / Vconv
            # Flush mode average velocity calculation:
            Pratio2 = Pmodin2 / Pmodout2
            C2=3/2*((Pratio2**2-1)/(Pratio2**3-1))
            Vmod2=pi*(dmod/2000)**2*Lmod
            Veloc2=Pref*Pconv*(Tj+273.15)*Lmod*Fmod2*Vconv*C2/(Pmodout2*Pconv*(Tref+273.15)*Vmod2)*100
            Flush=Veloc2*t

            result.append((Fill,Flush))
        #Function estimation:
        final=result[0]+result[1]
        if 12.3<=(final[0]+final[1])<=15.5:
            Elem1 =abs((final[0]+final[1])-15)
        elif (final[0]+final[1])<11.5 or (final[0]+final[1])>16:
            Elem1 = 3 * abs((final[0] + final[1]) - 15)
        else:
            Elem1 = 1.5 * abs((final[0]+final[1]) - 15)

        if 14.5<=(final[2]+final[3])<=17.5:
            Elem2 =abs((final[2]+final[3])-15)
        elif (final[2]+final[3])<14 or (final[2]+final[3])>21:
            Elem2 = 3 * abs((final[2]+final[3]) - 15)
        else:
            Elem2 = 1.5* abs((final[2]+final[3]) - 15)

        if (final[3]/final[2])<1:
            Elem3 =5
        elif (final[3]/final[2])<5:
            Elem3 =-((final[3]/final[2])-1)/2
        else:
            Elem3 = -2.5
        elem.append((Elem1,Elem2,Elem3))

        Funct=2*(Elem1+Elem2)+Elem3
        Elements.append(elem)
        FunctList.append(Funct)
        fullresult.append(result[0]+result[1])

    #Function minimum:
    print(np.array(fullresult))
    solution=min(FunctList)
    ind=FunctList.index(min(FunctList))


    #fill fields with results:
    entries['2D Flow (ml/min)'].delete(0, 'end')
    entries['2D Flow (ml/min)'].insert(0, new_array[ind][5])

    entries['Modulation period (s)'].delete(0, 'end')
    entries['Modulation period (s)'].insert(0, new_array[ind][6])

    entries['Injection time (s)'].delete(0, 'end')
    entries['Injection time (s)'].insert(0, new_array[ind][7])

    entries['Function'].delete(0, 'end')
    entries['Function'].insert(0, round(solution,2))
    if float(entries['Function'].get())<=10:
        entries['Function'].configure({"background": "Green"})
    elif float(entries['Function'].get()) < 20:
            entries['Function'].configure({"background": "Orange"})
    else:
        entries['Function'].configure({"background": "Red"})

    entries['Flush/Fill'].delete(0, 'end')
    entries['Flush/Fill'].insert(0, round(fullresult[ind][1]/fullresult[ind][0],2))
    if float(entries['Flush/Fill'].get())>=2.5:
        entries['Flush/Fill'].configure({"background": "Green"})
    elif float(entries['Flush/Fill'].get()) > 1 and float(entries['Flush/Fill'].get())<2.5:
            entries['Flush/Fill'].configure({"background": "Orange"})
    else:
        entries['Flush/Fill'].configure({"background": "Red"})




def makeform(root, fields):
    entries = {}
    for field in fields:
        if field == 'Function':
            #print(field)
            row = tk.Frame(root)
            lab = tk.Label(row, width=10, text=field + ": ", anchor='w',font='Helvetica 9 bold')
            ent = tk.Entry(row)
            ent.insert(0, "0")
            row.pack(side=tk.TOP,
                     fill=tk.X,
                     padx=50,
                     pady=5)
            lab.pack(side=tk.LEFT)
            ent.pack(side=tk.RIGHT,
                     expand=tk.YES,
                     fill=tk.X)
        elif field == 'Flush/Fill':
            #print(field)
            row = tk.Frame(root)
            lab = tk.Label(row, width=10, text=field + ": ", anchor='w',font='Helvetica 9 bold')
            ent = tk.Entry(row)
            ent.insert(0, "0")
            row.pack(side=tk.TOP,
                     fill=tk.X,
                     padx=50,
                     pady=5)
            lab.pack(side=tk.LEFT)
            ent.pack(side=tk.RIGHT,
                     expand=tk.YES,
                     fill=tk.X)

        elif field == '1D Length* (m)':
            #print(field)
            row = tk.Frame(root)
            lab = tk.Label(row, width=22, text=field + ": ", anchor='w')
            ent = tk.Entry(row)
            ent.insert(0, "30")
            row.pack(side=tk.TOP,
                     fill=tk.X,
                     padx=5,
                     pady=5)
            lab.pack(side=tk.LEFT)
            ent.pack(side=tk.RIGHT,
                     expand=tk.YES,
                     fill=tk.X)

        elif field == '2D Length* (m)':
            #print(field)
            row = tk.Frame(root)
            lab = tk.Label(row, width=22, text=field + ": ", anchor='w')
            ent = tk.Entry(row)
            ent.insert(0, "5")
            row.pack(side=tk.TOP,
                     fill=tk.X,
                     padx=5,
                     pady=5)
            lab.pack(side=tk.LEFT)
            ent.pack(side=tk.RIGHT,
                     expand=tk.YES,
                     fill=tk.X)

        elif field == '1D Diameter* (mm)':
            #print(field)
            row = tk.Frame(root)
            lab = tk.Label(row, width=22, text=field + ": ", anchor='w')
            ent = tk.Entry(row)
            ent.insert(0, "0.25")
            row.pack(side=tk.TOP,
                     fill=tk.X,
                     padx=5,
                     pady=5)
            lab.pack(side=tk.LEFT)
            ent.pack(side=tk.RIGHT,
                     expand=tk.YES,
                     fill=tk.X)

        elif field == '2D Diameter* (mm)':
            #print(field)
            row = tk.Frame(root)
            lab = tk.Label(row, width=22, text=field + ": ", anchor='w')
            ent = tk.Entry(row)
            ent.insert(0, "0.25")
            row.pack(side=tk.TOP,
                     fill=tk.X,
                     padx=5,
                     pady=5)
            lab.pack(side=tk.LEFT)
            ent.pack(side=tk.RIGHT,
                     expand=tk.YES,
                     fill=tk.X)

        elif field == 'Initial temperature* (°C)':
            #print(field)
            row = tk.Frame(root)
            lab = tk.Label(row, width=22, text=field + ": ", anchor='w')
            ent = tk.Entry(row)
            ent.insert(0, "70")
            row.pack(side=tk.TOP,
                     fill=tk.X,
                     padx=5,
                     pady=5)
            lab.pack(side=tk.LEFT)
            ent.pack(side=tk.RIGHT,
                     expand=tk.YES,
                     fill=tk.X)

        elif field == 'Final temperature* (°C)':
            #print(field)
            row = tk.Frame(root)
            lab = tk.Label(row, width=22, text=field + ": ", anchor='w')
            ent = tk.Entry(row)
            ent.insert(0, "300")
            row.pack(side=tk.TOP,
                     fill=tk.X,
                     padx=5,
                     pady=5)
            lab.pack(side=tk.LEFT)
            ent.pack(side=tk.RIGHT,
                     expand=tk.YES,
                     fill=tk.X)


        elif field == '2D outlet pressure* (psi)':
            #print(field)
            row = tk.Frame(root)
            lab = tk.Label(row, width=22, text=field + ": ", anchor='w')
            ent = tk.Entry(row)
            ent.insert(0, "14.696")
            row.pack(side=tk.TOP,
                     fill=tk.X,
                     padx=5,
                     pady=5)
            lab.pack(side=tk.LEFT)
            ent.pack(side=tk.RIGHT,
                     expand=tk.YES,
                     fill=tk.X)

        elif field == '1D Flow* (ml/min)':
            #print(field)
            row = tk.Frame(root)
            lab = tk.Label(row, width=22, text=field + ": ", anchor='w')
            ent = tk.Entry(row)
            ent.insert(0, "0.15")
            row.pack(side=tk.TOP,
                     fill=tk.X,
                     padx=5,
                     pady=5)
            lab.pack(side=tk.LEFT)
            ent.pack(side=tk.RIGHT,
                     expand=tk.YES,
                     fill=tk.X)
        elif field == '2D Flow (ml/min)':
            #print(field)
            row = tk.Frame(root)
            lab = tk.Label(row, width=22, text=field + ": ", anchor='w')
            ent = tk.Entry(row)
            ent.insert(0, "0")
            row.pack(side=tk.TOP,
                     fill=tk.X,
                     padx=5,
                     pady=5)
            lab.pack(side=tk.LEFT)
            ent.pack(side=tk.RIGHT,
                     expand=tk.YES,
                     fill=tk.X)
        elif field == 'Modulation period (s)':
            #print(field)
            row = tk.Frame(root)
            lab = tk.Label(row, width=22, text=field + ": ", anchor='w')
            ent = tk.Entry(row)
            ent.insert(0, "0")
            row.pack(side=tk.TOP,
                     fill=tk.X,
                     padx=5,
                     pady=5)
            lab.pack(side=tk.LEFT)
            ent.pack(side=tk.RIGHT,
                     expand=tk.YES,
                     fill=tk.X)

        elif field == 'Injection time (s)':
            #print(field)
            row = tk.Frame(root)
            lab = tk.Label(row, width=22, text=field + ": ", anchor='w')
            ent = tk.Entry(row)
            ent.insert(0, "0")
            row.pack(side=tk.TOP,
                     fill=tk.X,
                     padx=5,
                     pady=5)
            lab.pack(side=tk.LEFT)
            ent.pack(side=tk.RIGHT,
                     expand=tk.YES,
                     fill=tk.X)

        entries[field] = ent

    return entries


if __name__ == '__main__':
    root = tk.Tk()
    root.title('Flow modulation calculator')
    ents = makeform(root, fields)
    OptionList = [
        "Hydrogen",
        "Nitrogen",
        "Helium",
        "Argon"
    ]
    carrierGas = tk.StringVar(root)
    carrierGas.set(OptionList[0])
    opt = tk.OptionMenu(root, carrierGas, *OptionList)
    opt.config(width=30, font=('Helvetica', 8))
    opt.pack()

    b2 = tk.Button(root, text='Calculate',
           command=(lambda e=ents: calculate(e)))
    b2.pack(side=tk.LEFT, padx=5, pady=5)
    b3 = tk.Button(root, text='Quit', command=root.quit)
    b3.pack(side=tk.LEFT, padx=5, pady=5)
    root.mainloop()
```